\documentclass{aa}

\usepackage{natbib}
\bibpunct{(}{)}{;}{a}{}{,} 
\usepackage{graphics}   
\usepackage{graphicx}   
\usepackage{epstopdf}
\usepackage{longtable}   
\usepackage{url}      
\usepackage{bm}        
\usepackage[varg]{txfonts}
\usepackage{pdflscape}
\usepackage{supertabular}
\usepackage{hyperref}   
\usepackage{wasysym}

\hypersetup{
    bookmarks=true,         
    unicode=false,          
    colorlinks=true,       
    linkcolor=blue,          
    citecolor=blue,        
    filecolor=blue,      
    urlcolor=blue           
}
\usepackage{color}

\begin{document}

\title{Star-planet interactions} 
\subtitle{V. Dynamical and equilibrium tides in convective zones}

\author{Suvrat Rao\inst{1,2}, Georges Meynet\inst{1}, Patrick Eggenberger\inst{1},  Lionel Haemmerl\'e\inst{1}, Giovanni Privitera\inst{1}, Cyril Georgy\inst{1}, Sylvia Ekstr{\"o}m\inst{1}, and Christoph Mordasini\inst{3}
}

 \authorrunning{Rao et al.}

 \institute{Geneva Observatory, University of Geneva, Maillettes 51, CH-1290 Sauverny, Switzerland
\and Indian Institute of Technology, Kharagpur, India
\and 
Physikalisches Institut, University of Bern, CH-3012 Bern}

\date{Received /
Accepted}
\abstract{
When planets are formed from the protoplanetary disk and after the disk has dissipated, the evolution of their orbits is governed by tidal interactions, friction, and gravitational drag, and also by changes in the mass of the star and planet. 
These interactions may change the initial distribution of the distances between the planets and their host star
by expanding the original orbit, by contracting it (which may cause an engulfment of the planet by the star), or by destroying the planet.}
{We study the evolution of the orbit of a planet orbiting its host star under the effects of equilibrium tides, dynamical tides, drag (frictional and gravitational), and stellar mass loss.}
{We used the Geneva stellar evolution code to compute the evolution of stars with initial masses of 1 and 1.5 M$_\odot$ with different rotation rates at solar metallicity. The star is evolved from the pre-main-sequence (PMS) up to the tip of the red giant branch. We used these models as input for computing the evolution of the planetary orbits. We
explored the effects of changing the planet masses (of 1 Earth mass up to 20 Jupiter masses), the distance between the planet and the star (of 0.015 and more than 3 au), 
the mass, and the spin of the star. We present results when only the equilibrium tide was accounted for and when both equilibrium and dynamical tides were accounted for. The expression for the dynamical tide is a frequency-averaged dissipation of tidally excited inertial waves, obtained from a piecewise homogeneous two-layer stellar model. Gravity wave damping was neglected. 
}
{Dynamical tides in convective zones have a significant effect on planetary orbits only during the PMS phase and only for fast-rotating stars. 
They have no significant effects during the PMS phase for initially slow-rotating stars and during the red giant branch phase, regardless of the initial rotation. 
In the plots of initial orbital distance versus planetary mass, we show the regions that lead to engulfment or any significant changes in the orbit. 
As a result of orbital evolution,  a region near the star can become devoid of planets after the PMS phase. We call this zone the {\it \textup{planet desert}}, and its extent depends  sensitively on stellar rotation.
An examination of the  planet distribution as a function of distance to the host star and mass can provide constraints on current computations.}
{}

\maketitle

\titlerunning{Impact of dynamical tides}
\authorrunning{Rao et al.}

\section{Introduction}

Studying the evolution of the orbits of planets after their formation is of great interest. Some reasons for this are as follows:
\begin{itemize}
\item First, it allows us to explore the links between an observed configuration of a planetary system at a given time with the initial configuration, that is, its configuration at birth. 
\item Second, the evolution of the planetary orbits affects some observable properties of their host star \citep{livio84, soker84, sackmann93, Rasio1996, siess99I, siess99II, villaver07, sato08, villaver09, carlberg09, nordhaus10, kunimoto11, bear11, mustill12, nordhaus13, villaver14, GIOI, GIOII, GIOIII, GIOIV}. Typically, the tidal interactions between star and planet cause a transfer of angular momentum between the planetary orbital angular momentum and stellar spin 
angular momentum. This results in a change in the spin angular momentum (and therefore the surface velocity) of the star.
These changes also affect the tides through modifications of the corotation radius (radius of a circular orbit such that the orbital period is equal to the stellar spin period) and of the amplitude of the tides.
A more dramatic
consequence of these tidal interactions can be the engulfment of the planet by the star, causing a further change in its angular momentum and an increase in the abundances of some elements in its envelope, such as lithium.
\item Third, the distributions of planets as a function of their distance to the host star evolve with time as a result of these interactions. Such a feature could be compared with observations when
a sufficient amount of data is gathered, through sampling the masses of the planets and stars, the stellar rotation and ages, and the metallicity.
\end{itemize}
The physics behind the tidal interactions is complex and not fully understood. Tides are generated by differential gravitational forces inside the body feeling the tides (here, the star) owing to the presence of a nearby companion (the planet). Differential gravity leads to some changes in mass distribution inside the star. Since the planet orbits the star, these changes evolve as a function of time. This dynamic mass-distribution variation 
in turn modifies the gravitational field exerted by the star  and thus affects the orbit of the generator of the tides. 
This is a complex loop, resulting in the generation of waves inside the star and in modifying the planetary orbits. 

Hopefully, however, observations will help constrain this physics. In this context, it would be 
very interesting to observe a planet whose orbit
evolves rapidly as a function of time owing to tides. This would produce an observable change in stellar spin and in orbital period as a function of time that might give clues on the strength of the tides, provided sufficient information about the
peculiar star-planet system is available. Some such attempts have been made previously \citep{Birkby2014, Wilkins2017, Patra2017}. 
Detecting such an effect for planets around red giants might be an interesting scientific objective \citep{GIOIV}. This might be possible with PLATO \citep{Rauer2014}.

Another approach is
to study from theory the consequences  of tidal interactions on observable properties of stars.
An example question is whether the fraction of fast-rotating red giants is compatible with the expected fraction of red giants that engulf planets. 
To make progress in this direction, we need to determine whether planets can deviate from their original orbit by being kicked out or through engulfment by the host star
during the phases that precede the red giant branch (RGB) phase.

In previous works of our team, we studied the evolution of the orbits of planets with masses between 1 and 15 M$_{\rm \jupiter}$ (where M$_{\rm \jupiter}$ is the mass of Jupiter), orbiting stars with masses between 1.5 and 2.5 M$_\odot$ at initial distances on the zero-age main-sequence (ZAMS) between 0.5 and 1.5 au. We considered only the case of a star with metallicity $Z$=0.020, and only equilibrium tides were considered. Our computations assumed that tides only become important when a convective envelope appears, and we started our computations from the ZAMS. \citet{GIOI} showed that the high surface velocity of some red giants cannot be obtained without any interaction with a close companion. We also verified that if such an interaction occurs with a planet, reasonable conditions exist to allow the star to reach the high surface velocities that are observed. In a second paper, we followed the evolution of the red giant after an engulfment to study how its surface velocity evolves as a function of time. We also studied the impact of a planet engulfment on changes of the red giant surface abundances   \citep{GIOII}. 
The fast rotation acquired by a red giant after the engulfment of a planet may trigger a dynamo and the appearance of strong surface magnetic fields. We studied this possibility in \citet{GIOIII}. Finally, we also investigated the possibility of
detecting changes in planetary orbits around red giants that are due to tides by observing transiting planets \citep{GIOIV}. 

In this work, we wish to go beyond these works, first by considering the effect of the dynamical tide in external convective zones (see Sect.~2) in addition to the equilibrium tide.
Moreover, we begin our computation from the PMS phase, much earlier than the ZAMS stage, where important convective envelopes are present and thus tidal forces can be strong.
The impact of dynamical tides in the convective zone has recently been studied by \citet{Bol2016} and \citet{Gal2017}
for stars with masses between 0.3 and 1.4 M$_\odot$. The effects of a change in metallicity have been studied by \citet{Bol2017}. 
Some aspects that distinguish this work from the previous papers are listed below.
\begin{itemize}
\item We use an equation for the evolution of the orbit that also accounts
for stellar mass loss and for the frictional and gravitational drag in addition to the tides  (see Eq.~9 below). 
\item The effect of equilibrium tides is accounted for in a comprehensive way. In previous works that studied the effect of dynamical tides in convective zones,  a constant dissipation of the equilibrium tide was assumed (i.e., a constant value for the equilibrium tide dissipation factor $\sigma_\star$). We here follow the change in efficiency of the equilibrium tides when the stellar properties change during the evolution. This is particularly important during the PMS and red giant phases, when the convective envelope is developed and evolves rapidly (see below).
\item Our study covers the whole RGB phase.
\end{itemize}
An important first objective is to determine the extent to which the inclusion of the dynamical term affects results that were previously obtained
concerning the engulfment of planets during the RGB. We therefore focus on stars with masses equal
to 1 and 1.5 M$_\odot$. 
Second, we study the PMS phase and the role of the dynamical tide during that phase for planet masses of between
0.1 and 20 Jupiter masses (even to an Earth mass planet in the case of 1 M$_\odot$), and for initial distances between the planet and the star from 0.015 au up to 4 au.
Thus, we significantly enlarge the parameter space studied in our previous works. 

In Sect.~2 we present the equations we used to compute the evolution of the orbits and our method. We briefly discuss some general aspects of dynamical and equilibrium 
tides in Sect.~3.
Numerical computations of orbital evolutions are commented on in Sect.~4, and the fates of planets of different masses at various distances from their parent stars are described in Sect.~5. 
Finally,  the main
conclusions and some limitations of our approach are discussed in Sect.~6.

\section{Physics of our computations}\label{sec:2}

\subsection{Expression for the equilibrium tide}

\begin{figure*}
\centering
\includegraphics[width=.5\textwidth, angle=0]{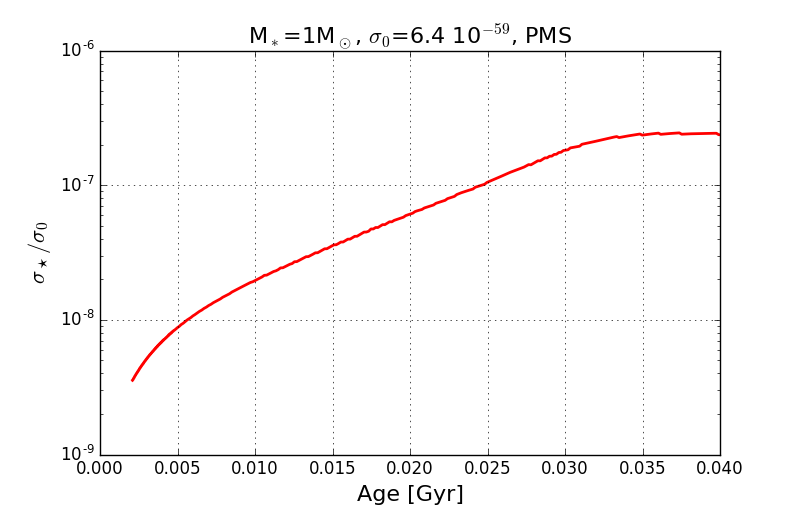}\includegraphics[width=.5\textwidth, angle=0]{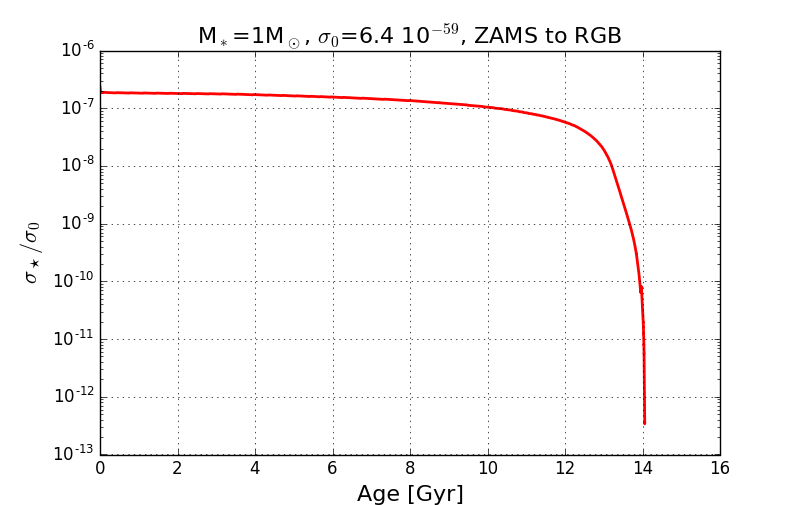}
\caption{Evolution of $\sigma_\star$ as a function of time as given by Eq.~(4). 
We consider the case where $P_{\rm orb}$ is equal to one day.  $\sigma_0$ is the normalization as in \citet{Hansen2012}: $\sigma_0=6.4$ $10^{-59}$ g$^{-1}$ cm$^{-2}$ s$^{-1}$.
{\it Left panel:}  During the PMS phase.
{\it Right panel:} From the ZAMS up to the tip of the RGB.
}
\label{sigma}
\end{figure*}

The expression of the equilibrium tide is based on the work by 
\citet{Hut1981}. It has been adapted for use in the frame of close binary star evolution by  \citet{Hurley2002} and in the computation of planet orbits by \citet{Rasio1996}.
The equilibrium tide \citep[see][]{zahn66,alexander76,zahn77,zahn89,livio_soker84,villaver09,mustill12,villaver14} is accounted for only when an external convective zone is present. Its expression is taken as in \citet{GIOI}. For self-consistency, we recall the expression below. Assuming a circular orbit in the equatorial plane of the star, the evolution of the radius of the orbit, $a$, due to equilibrium tide, is given by
\begin{equation}
(\dot{a}/a)_{\rm eq}=\frac{f}{\tau}\frac{M_{\rm env}}{M_{\star}}q(1+q) \left( \frac{R_{\star}}{a}\right)^{8}\left( \frac{\Omega_{\star}}{\omega_{\rm pl}}-1\right),
\end{equation}
where $f$ is a numerical factor (see below),
$M_{\rm env}$ is the mass of the convective envelope, $q=M_{\rm pl}/M_{\star}$, with $M_{\rm pl}$ the mass of the planet, and $M_{\star}$ that of the star,
$\Omega_{\star}$ is the angular velocity at the surface of the star, 
$\omega_{\rm pl}=2\pi/P_{\rm orb}$ is the orbital angular velocity of the planet, and $\tau$ is the eddy turnover timescale \citep{villaver09},
\begin{equation}
\tau=\left[\frac{M_{\rm env}(R_{\star}-R_{\rm env})^{2}}{3L_{\star}} \right]^{1/3}\ \ \ \ ,
\end{equation}
with $R_{\rm env}$ being the radius at the base of the convective envelope of the star and $L_{\star}$ the luminosity of the star. 
 This expression is slightly different from the one given by \citep{Rasio1996} in which
the term $(R_{\star}-R_{\rm env})^{2}$ is replaced by $R_{\rm env}(R_{\star}-R_{\rm env})$.
The factor $f$ is equal to 1 except when
$\tau > P_{\rm orb}/2$; in that case, it becomes \citep{Goldreich1977}
$$f=\left({P_{\rm orb}\over 2\tau}\right)^2.$$
Depending on whether the planet is beyond or inside the corotation radius, which is 
the distance at which the orbital period of the planet is equal to the stellar spin period ($a_{\rm corot}=(G (M_\star+M_{\rm pl})/\Omega_\star^2)^{1/3}$), tides may cause the distance between the star and the planet to increase or to decrease.
This aspect is included in Eq.~(1) through the sign of $\left( \frac{\Omega_{\star}}{\omega_{\rm pl}}-1\right)$.

The expression for the equilibrium tide used by \citet{Bol2016}, \citet{Gal2017}, and \citet{Bol2017} is given by
\begin{equation}
(\dot{a}/a)_{\rm eq}=9 \sigma_\star  M_\star R_{\star}^2 q(1+q) \left( \frac{R_{\star}}{a}\right)^{8}\left( \frac{\Omega_{\star}}{\omega_{\rm pl}}-1\right)\ \ \ \ ,
\end{equation}
where $\sigma_\star$ is the equilibrium tide dissipation factor. In
\citet{Bol2016}, \citet{Gal2017}, and \citet{Bol2017} a constant value of $\sigma_\star$ is assumed for a given initial mass star. As noted by \citet{Gal2017},
such an assumption constitutes a simplified model. To study this in more detail, we first express
$\sigma_\star$ by comparing Eqs. (1) and (3):
\begin{equation}
\sigma_\star= \frac{1}{9 M_{\star} R_\star^2} \frac{f}{\tau}\frac{M_{\rm env}}{M_{\star}}.
\end{equation}
In Appendix A, we explain that starting from the general expression defining $\sigma_\star$ \citep{Eggleton1998} and using
the mixing length theory, it is possible to derive Eq. (4) above. Figure~\ref{sigma} shows the evolution of $\sigma_\star$
as a function of time for the PMS phase (left panel) and from the ZAMS until the tip of the red giant branch (RGB) phase (right panel) for a 1 M$_\odot$ star and with a forcing period of one day.
A constant value of $\sigma_\star$ is certainly a fair approximation for the main-sequence phase, but
it is a poor approximation in phases when the convective envelope is more developed and evolves more rapidly than during the main-sequence phase, as is the case during the PMS and red giant phases, for instance.




Moreover,
a constant value would mean that only a given orbital period (or a given distance of the planet to its host star) is considered. 
For close-in planets, for which the turnover time is longer than the orbital period divided by 2, $f$ is indeed not constant and depends
on $P_{\rm orb}$. For a 1 M$_\odot$ star, that is, for the case shown in Fig.~\ref{sigma},
$\tau$ is always longer than 10 days (see Fig.~\ref{struc}), and thus longer than 
the orbital period (here 1 day, this corresponds to a distance of 0.02 au.) divided by 2.
Therefore, $f$ depends on
$P_{\rm orb}$. For an orbital period
twice as long, for instance (2 days, corresponding to a distance of 0.03 au), the curve for $\sigma_\star$ would be shifted upward by a factor 4. 

We here used Eq.~(1) for the equilibrium tide, as described, and thus by construction a non-constant value for $\sigma_\star$. This allows us
to account for the changes of the global stellar properties, the mass of the convective envelope, and the convective turnover time 
with the evolution of the star, an aspect that is crucial especially for the phases during which the magnitudes of tides are large.

\subsection{Expression for the dynamical tide}

Dynamical tides in convective zones occur through inertial waves, that is,  waves inside the star whose restoring force
is the Coriolis force. Waves are also excited at the base of the convective zone; these are known as Hough waves. They propagate into the radiative zone,  and if they are not coherently reflected off the center of the star, they dissipate energy and thus contribute to the exchange of angular momentum between the planet orbit and the star \citep[see more details in][] {Goodman1998, Ogilvie2007, Barker2010}. These dynamical tides in radiative zones are not considered here. 
We briefly discuss this point in Sect.~6.

The dynamical tide is accounted for only when there is an external convective zone and  when $\omega_{\rm pl} < 2\Omega_{\star}$. 
 Its expression is given by \citep{Ogil2013, Mathis2015}
\begin{equation}
(\dot{a}/a)_{\rm dy}=\left(9 \over {2 Q_{\rm prime,d}} \right) q \omega_{\rm pl} \left({R_{\star} \over a}\right)^{5} {(\Omega_{\star}-\omega_{\rm pl}) \over |\Omega_{\star}-\omega_{\rm pl}|}\ \ \ \ ,
\end{equation}
with $Q_{\rm prime,d}=3/(2 D_{w})$, and $D_{w}=D_{0w} D_{1w} D_{2w}^{-2}$, with
\begin{equation}
D_{0w}=
{100\pi \over 63} 
\epsilon^2
{\alpha^5 \over 1-\alpha^5}
( 1- \gamma)^2
\end{equation}
\begin{equation}
D_{1w}=(1-\alpha)^4 \left(1+2\alpha+3\alpha^2+{3 \over 2}\alpha^3\right)^2 \left(1+{1-\gamma \over \gamma} \alpha^3\right)
\end{equation}
\begin{equation}
D_{2w}=1+{3\over 2}\gamma +{5 \over 2\gamma}\left(1+{\gamma \over 2}-{3\gamma^2 \over 2}\right) \alpha^3-{9\over 4}(1-\gamma)\alpha^5 
,\end{equation}
where $\alpha=R_c/R_\star$,  $\beta=M_c/M_\star$, $\gamma={\alpha^3(1-\beta)\over \beta(1-\alpha^3)}$ (ratio of the densities of the envelope to the core in the two-layer model used to deduce the expression of the dynamical tide), $\epsilon={\Omega_{\star} \over \sqrt{G M_\star \over R_\star^3}}$. $M_c$ and $R_c$
are the mass and radius of the radiative core, respectively.

In the derivation of the expressions indicated above for the dynamical tides in convective zones, certain simplifications have been made in order to obtain an analytical result for the frequency-averaged dissipation rate.
We briefly discuss some of these limitations in the conclusions.

We account for equilibrium tides at all times, regardless of whether the dynamical tides are active,
although when they are active and have a significant impact on the orbit, dynamical tides are much stronger than the equilibrium tides. 
Hence, in our case, the results would be unaffected by neglecting equilibrium tides in situations where dynamical tides are active. 


\subsection{Equation for the orbit evolution}

For simplicity, we assume that the planet follows a circular orbit with no obliquity, as in our previous papers.
The orbit is not only affected by tides, but also by other factors such as changes in the masses of the star and the planet, as well as by frictional and gravitational drags that
reflect the interaction of the planet with the interplanetary medium \citep[see][]{zahn66, alexander76, zahn77, zahn89, livio_soker84, villaver09, mustill12, villaver14}.  The
equation describing the total change of $a$ is given by
\begin{equation}
\left(\frac{\dot{a}}{a} \right)=
-\frac{\dot{M}_{\star}+\dot{M}_{\rm pl}}{M_{\star}+M_{\rm pl}}
-\frac{2}{M_{\rm pl}v_{\rm pl}}\left[F_{\rm fri} + F_{\rm gra}\right]
+\left(\frac{\dot{a}}{a} \right)_{t}
\label{equa:evoorb}
,\end{equation}
where $\dot{M}_{\star}=-\dot{M}_{\rm loss}$ , with $\dot{M}_{\rm loss}$ being the mass-loss rate (here given as a positive quantity). $M_{\rm pl}$ and $\dot{M}_{\rm pl}$ are the planetary mass and the rate of change in the planetary mass, $v_{\rm pl}$ is the velocity of the planet. $F_{\rm fri}$ and $F_{\rm gra}$ are the frictional and gravitational drag forces, respectively, while $(\dot{a}/a)_{t}$ is the term that takes into account the effects of the equilibrium and dynamical tidal forces whose expressions are given above. The expressions for the
frictional and gravitational drags were taken as in \citet{villaver09}, \citet{mustill12}, and \citet{villaver14}.
Frictional drag occurs because the planet is subject to some braking while moving in the interplanetary medium. The density of this medium can be high enough when the star loses a significant amount of mass by stellar winds.
This term causes a decrease in the radius of the planetary orbit and does not imply any change in the stellar angular momentum. The gravitational drag arises because the movement of the planet along its orbit produces a wake in the uniform gaseous medium. A drag force results from the gravitational attraction between the planet and its wake. This term, as the frictional drag,
causes a decrease in the radius of the orbit but does not change the stellar angular momentum.

We did not account for the effects of stellar magnetic fields
as explored in \citet{GIOIII}, nor did we account for possible interactions between the stellar and planetary magnetic field as in \citet{Stru2017}. 
The evolution of the spin of the model with one solar mass accounts for the wind-magnetic braking process according to the law of \citet{Krish1997}.
Effects of stellar mass loss on the planet orbit through the drag forces are considered only during the RGB phase.

We did not consider the effect of planet evaporation (thus we assume $\dot{M}_{\rm pl}$=0), nor did we consider
the possibility that the planet might be disrupted by tides below the Roche limit. All these aspects involve physics
describing the planet itself (e.g., the Roche limit depends on the mean density of the planet, typically the Roche limit for Jupiter and the Sun  is 0.01 au). 
These points will be studied in a forthcoming paper.

\subsection{Stellar models}

The equation describing the evolution of the orbit (see Eqs.~1, 5, and 9) shows that various quantities resulting from rotating stellar models are required. For this purpose, we computed rotating stellar models for stars with 1 and 1.5 M$_\odot$ at solar metallicity.
These models were computed with the Geneva stellar evolution code \citep{egg08}, using the same physical ingredients as in \citet{ekstrom12}. 
The required stellar inputs for computing the evolution of the orbit during the PMS and the RGB phases are given in a graphical form in the appendix. 

We used these inputs to compute the orbit evolution.
The evolution of the rotation of the convective envelope was corrected for by taking into account 
the orbital changes that are due to the tides (we note that frictional and gravitational drag forces do not modify the stellar angular momentum). To do this, we computed the changes in orbital angular momentum that are due to tides, and we modified the angular velocity of the convective envelope of the star
by removing (when the planet orbits the star beyond the corotation radius) or adding (when the planet orbit is inside the corotation radius) this angular momentum to the angular momentum of the convective envelope. The convective envelope is assumed to rotate as a solid body. This method assumes first 
that these changes will not affect the other quantities describing the star, such as  the mass of the convective envelope or the stellar mass-loss rate. Second, the method assumes that the angular momentum in the convective envelope has no time to be transported 
inside the star by processes such as shear instabilities or meridional currents. 
When strong exchanges of angular momentum occurred between the orbital angular momentum and spin angular momentum, we limited our calculations to phases during which an external convective zone exists. 
To go beyond these phases would require coupling the stellar evolution code and the orbital evolution more tightly, as we did in \citet{GIOI}, in order to simultaneously follow the evolution of the star and of the orbit. 

\section{Equilibrium versus dynamical tides in convective envelopes}\label{sec:1}

In this section, we briefly discuss the conditions that favor one of the two tidal forces during the PMS and the RGB phases.
As is obvious from Eq.~(1), the term $(R_\star/a)^8$ has a dominant role in the equilibrium tide expression. Since the radii of stars are one to two orders of magnitude larger
during the RGB phase, everything else being equal, this causes the equilibrium tides to be much stronger during the RGB phase than during the PMS phase.
Figure~\ref{struc} shows that the other factors, such as the turnover time and the ratio of the mass of the envelope to the mass of the star, are similar during both phases.


The dynamical term also shows a strong dependence on $R_\star/a$, although it is not as strong as in the equilibrium tide, indicating that everything else being equal, the increase
due to this term is not as significant as it is for the equilibrium tide when the stellar radius expands. 
The dynamical term, through its dependence on Q$_{\rm prime}$, is proportional to $\epsilon^2 \alpha^5/(1-\alpha^5)$.
Both $\epsilon$ and $\alpha$ are much larger during the PMS than during the RGB phase, making the tidal term more effective during the PMS than during the RGB phase.
The dependence on $\epsilon^2$ especially shows that dynamical tides are important for fast-rotating stars. 
Red giants are slow rotators (at least before any acceleration due to an interaction with a planet) 
and thus are much less favorable for showing a strong effect of dynamical tides.
From the considerations above, we expect that the equilibrium tide is more significant during the RGB phase than during the PMS, and that the inverse is true for the dynamical tide.

At a given phase, which tide will dominate? Considering a given orbital period (i.e.,\textup{} a given $a$), and keeping only the most important terms, we have

\begin{equation}
{
(\dot{a}/a)_{\rm eq}
\over (\dot{a}/a)_{\rm dyn}
}
\propto 
{
\left({R_\star \over R_\odot}\right)^3 Q_{\rm prime,d}.
}
\end{equation}

This ratio increases when the star evolves from the PMS to the RGB phase ($(R_\star/R_\odot)^3 Q_{\rm prime,d}$  is on the order of 10$^{4}$ during the PMS of our 1 M$_\odot$ model and passes from
10$^{6}$ at the beginning of the RGB to 10$^{17}$ at the tip of the RGB).
Thus we expect the equilibrium tide to become increasingly more significant than
the dynamical tide as the star evolves in time. This means that if the equilibrium tide is dominant during the PMS phase, it will continue to be dominant 
during the whole evolution up to the tip
of the RGB phase. When the dynamical tide dominates during the PMS, then nothing can be said from such considerations, and the complete expression has to be estimated. This is done in the next section.

\section{Evolution of the orbits}\label{sec:2}

\subsection{Pre-main-sequence phase}

\begin{figure*}
\centering
\includegraphics[width=.52\textwidth, angle=0]{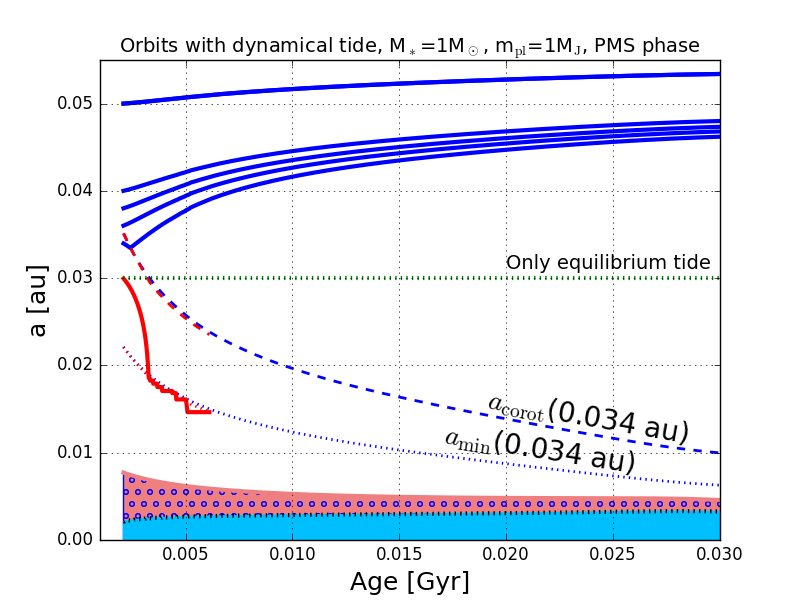}\includegraphics[width=.52\textwidth, angle=0]{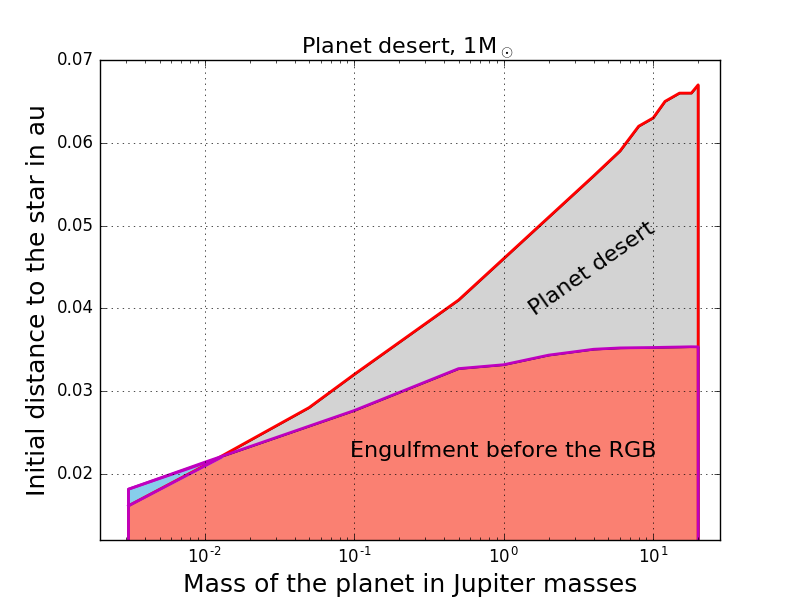}
\caption{{\it Left panel:} Evolution as a function of time of the orbits of Jupiter mass planets around a 1 M$_\odot$ star (continuous lines). 
Different initial distances to the star are considered. Blue lines show cases where tides tend to widen the orbits, and the red line shows a case
where the orbit shrinks. Only the beginning of the evolution is shown (see text). The blue dashed  and dotted lines show the corotation  ($a_{\rm corot}$) and the critical distance below which dynamical tides are no longer active ($a_{\rm min}$), respectively. 
The blue dashed and dotted lines correspond to the case when the initial distance is 0.034 au. The red dashed and dotted lines have the same meaning as the blue dashed and dotted lines, but correspond to an initial distance of 0.030 au. The green dotted line is the orbit for a planet starting at 0.03 au where the dynamical tide is neglected. The upper limit of the pink zone with blue circles corresponds to the radius
of the star in au. Its lower limit shows the radius at the base of the stellar convective envelope. The blue zone indicates the size of the stellar radiative interior.
{\it Right panel:} The red zone shows the region in the initial distance vs. planet mass plane where an engulfment occurs during the PMS phase. The upper limit of the light gray zone shows the minimum distance at which a planet beginning its evolution at a distance inside the light gray zone is moved by tides.  This gray zone is thus devoid of planets, not because of engulfment but because the orbit expands. The blue region
may contain planets. Planets can be present in that region only if they have begun their evolution above the blue zone and {\it \textup{have evolved}} there because of a net shrinking of their orbit (see text). Planets that have an {\it \textup{initial}} distance inside the blue
region are engulfed.
}
\label{P001z20S1PMS1aVSage1JUP}
\end{figure*}

The evolutions of a few orbits for a planet with one Jupiter mass around  a star of solar mass during the PMS phase are shown in the left panel of Fig.~\ref{P001z20S1PMS1aVSage1JUP}.
We assumed here a short formation process for the planet, since we began this computation only 2 Myr after the beginning of the PMS phase. 
Below we investigate the
effect of changing the time of the start of the computation.

We consider, for instance, the orbit beginning at an initial distance of 0.04 au. This orbit is beyond the corotation radius (the blue dashed line). As indicated above, for dynamical tides to be active, it is required that $\omega_{\rm pl} < 2\Omega_\star$. Using the relation $\omega_{\rm pl}=\sqrt{G (M_\star+M_{\rm pl})/a^3}$, where $G$ is the gravitational constant, the condition $\omega_{\rm pl} < 2\Omega_\star$ can be translated into a condition on the distance between the planet and the star. For dynamical tides to be active, it is required that $a > a_{\rm min}=(G (M_\star+M_{\rm pl})/(4\Omega_\star^2))^{1/3}$. The dotted line in the left panel of Fig.~\ref{P001z20S1PMS1aVSage1JUP} shows $a_{\rm min}$. The evolution of this line and the line showing the corotation $a_{\rm corot}$ depend on the evolution of the stellar rotation and thus are different for each star-planet system considered here where the stellar spin is changed due to tides. The blue (red) dotted line corresponds to the case  where the initial distance of the planet is 0.034 au (0.030 au). The corotation radii for all cases starting at initial distances above this limit are equal to the line for the case at 0.034 au, since for the larger distance cases shown in the figure, the changes in stellar spin that are due to tides are negligible.
The orbit here is beyond $a_{\rm min}$ and $a_{\rm corot}$, so that dynamical tides
are active and tend to widen the orbit. 

Dynamical tides clearly dominate equilibrium tides. The computation of an orbit without the dynamical tide starting at 0.03 au results in the horizontal dotted green line. Hence equilibrium tides have a negligible effect here.

Considering an orbit starting at 0.034 au (lower continuous blue curve), we observe a kick in the orbit. First the orbit shrinks, and then it widens. The orbit begins below the corotation radius and thus
shrinks under the effect of tides.  Conversely, when the orbit crosses the corotation radius, tides widen the orbit. Therefore the orbit bounces back on the corotation limit. This behavior has been found
by  \citet{Bol2016}, \citet{Gal2017}, and \citet{Bol2017}. 

We now consider the orbit beginning at an initial distance of 0.03 au (the red continuous curve). 
Since the planet is closer to its star, tides are stronger and 
prominently affect the orbit, which shrinks rapidly.
We note that the curve corresponding to the corotation radius (the red dashed curve) lies below the blue dashed curve because the spin of the star is very slightly accelerated by the tidal interaction with the planet.  Higher stellar spin shifts the corotation radius downward. 
The curve for $a_{\rm min}$ (the red dotted curve) is also slightly shifted downward (but this shift is too small to be clearly visible in the figure). At a certain time, the orbit passes below $a_{\rm min}$, which switches the convective dynamical term off. Here we did not consider dynamical tides in radiative zones, therefore dynamical tides disappear. Since the equilibrium tide is smaller than the dynamical tide at this stage, the shrinkage of the orbit slows down. The orbit then more or less follows the curve for $a_{\rm min}$. It will continue to do so until either the planet evaporates, is tidally destroyed, or is engulfed by the star.
In our simulation, we stopped the computation at an age of about 6.1 Myr because it requires very small time steps to precisely compute the evolution beyond this point.

For each planet mass, we can determine the initial distances below which an evolution typical of the red continuous curve in the left panel of Fig.~\ref{P001z20S1PMS1aVSage1JUP} occurs.
This is shown in the right panel of Fig.~\ref{P001z20S1PMS1aVSage1JUP} (see the red zone). The light gray zone labeled \textup{}"planet desert" corresponds to a region where
planets are kicked off  from their original orbit into wider orbits.

The small blue region shows what happens
to planets beginning their evolution just above the upper limit of the blue region. The orbits of these planets show a kick similar to the kick shown by the orbit for 0.034 au in the left panel of Fig.~\ref{P001z20S1PMS1aVSage1JUP},
but the shrinking part of the bounce is larger than the widening part, and thus the orbit has a smaller radius than the initial radius at the end of the kick, even smaller than the initial radius needed for an engulfment. 
The final distance of these planets would be in the blue zone. They survive at these small distances because, when they reach their final orbit, tides are less important because the mass of the convective envelope decreases.
We call this the 'survival zone'.

As can be deduced from the discussion above, the dynamical tide opens new channels for the orbital evolutions. 
This concerns planets orbiting their host stars at distances well below 0.1 au.

\subsubsection{Effect of changing the starting time}

As mentioned above, we can consider what happens when we begin to compute the orbit of a planet before or after the time considered in Fig.~\ref{P001z20S1PMS1aVSage1JUP} (i.e., 2 Myr). 
Different starting times would physically correspond to different durations for the formation of the planets and dissipation of the protoplanetary disk.

Changes in the orbits for Earth-like planets, for 1 Jupiter mass planets and 20 Jupiter mass planets for different starting times and distances are shown in Fig.~\ref{M0031TDIFF}. As explained above, the orbits of the planets falling into the star are expected to follow the $a_{\rm min}$ curve. We kept the orbit as given by our code, however, because as explained above, these planets will likely be evaporated or engulfed at a later time.

For Earth-mass planets, little difference is observed for those beginning at a distance of 0.02 au for different starting times. At a distance of 0.018 au, planets beginning their evolution at 1 and 2 Myr have a significantly shrunken orbit compared to those beginning at 5 Myr.
These planets would graze the surface of the star and would likely suffer complete evaporation or at least be eroded and later engulfed.  
In conclusion, changing the starting time causes a small shift in the limits of the various zones shown in the right panel of Fig,~\ref{P001z20S1PMS1aVSage1JUP}.
Globally shifting the starting time to higher values shifts the upper limits of the red and light gray zones downward.

This is even clearer in the middle and right panels of Fig.~\ref{M0031TDIFF}, which show that delaying the starting time for more massive planets decreases the limit for bouncing orbits as well as the limit for an engulfment during the PMS phase. Thus, the greater the time required for the formation of a planet is, the larger are the chances of its survival beyond the PMS phase. The physical interpretation of this observation is apparent, since the later into the PMS phase we begin the orbital evolution, the smaller the initial stellar convective zone, which would reduce the initial impact of the dynamical tides. Furthermore, the total duration for which the dynamical tides act would be shorter. Thus, the overall impact of dynamical tides during the PMS phase would be weaker, which in turn would lower the upper limits of the red and light gray zones.

\subsubsection{Effect of changing the planet mass}

\begin{figure*}
\centering
\includegraphics[width=.34\textwidth, angle=0]{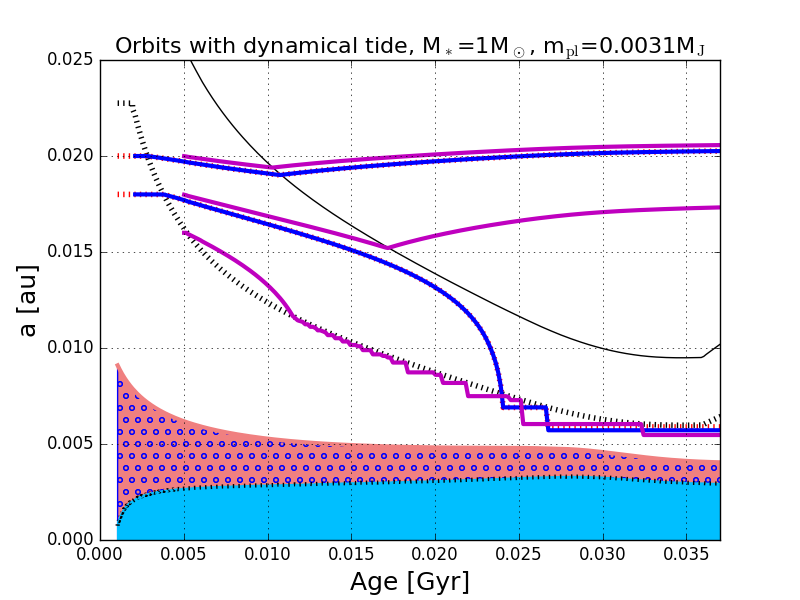}\includegraphics[width=.34\textwidth, angle=0]{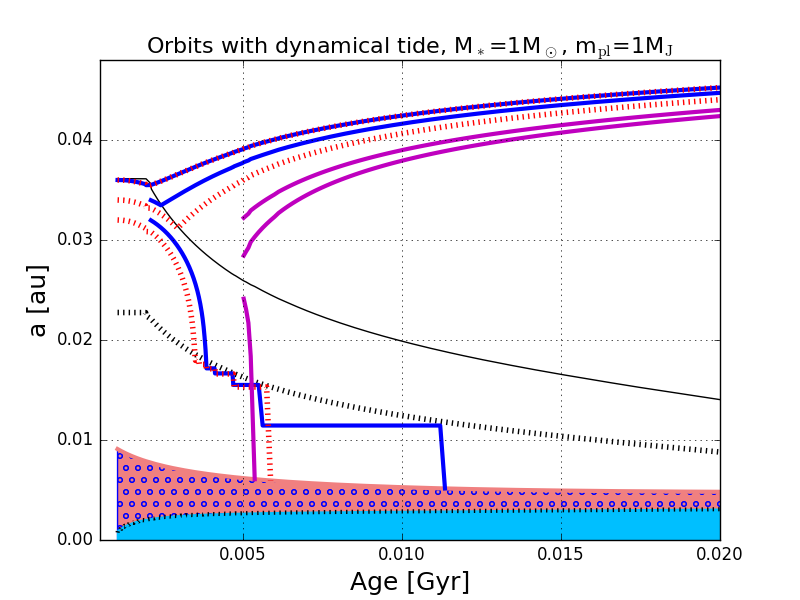}\includegraphics[width=.34\textwidth, angle=0]{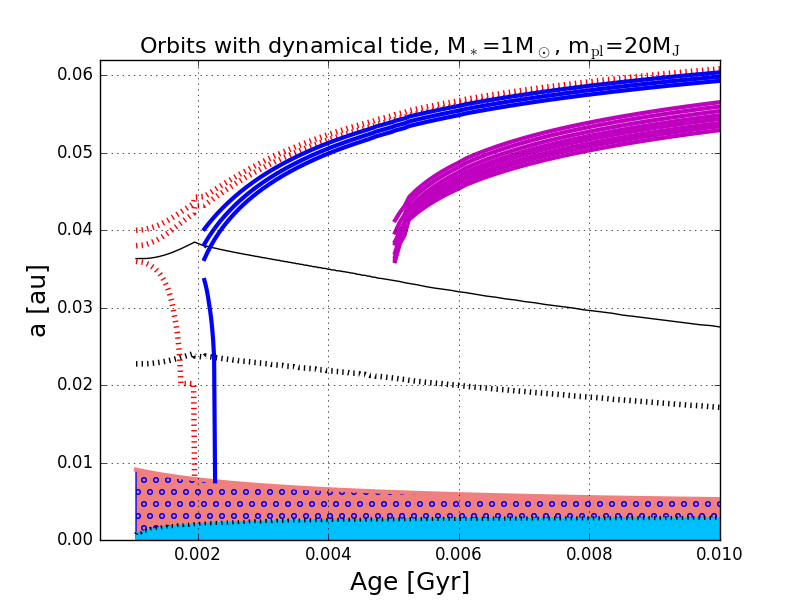}
\caption{Evolution of the orbits of 0.0031 (Earth-mass planet), 1 and 20 Jupiter mass planets around a 1 M$_\odot$ star, starting their orbital evolution at different times and distances to their host star. The red dotted lines show planets beginning their evolution at 1 Myr, the continuous blue line shows planets at 2 Myr, and the continuous magenta line shows planets at 5 Myr. The black continuous line depicts the corotation radius, and the black dotted line represents $a_{\rm min}$. These quantities correspond to cases where the starting time is 1 Myr. For the 0.0031 (one Earth-mass planet), 1 and 20 Jupiter mass planet,
they correspond to initial distances equal  to 0.1, 0.036, and 0.04 au, respectively.
The colored regions have the same meaning as in the left panel of Fig.~\ref{P001z20S1PMS1aVSage1JUP}.}
\label{M0031TDIFF}
\end{figure*}

As expected, more massive planets have larger engulfment zones than lighter planets (see Fig.~\ref{M0031TDIFF}). 
In the region where a planet is kicked off from its original orbit, the kick is also more significant when the mass of the planet is higher.
This is a consequence of increasing tides when the mass of the planet increases.

\subsubsection{Effect of changing the stellar mass}

\begin{figure*}
\centering
\includegraphics[width=.52\textwidth, angle=0]{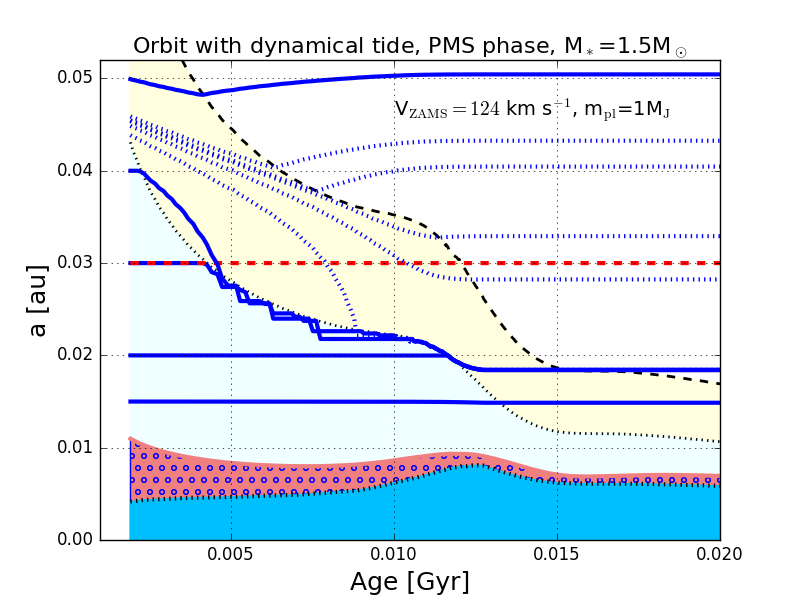}\includegraphics[width=.52\textwidth, angle=0]{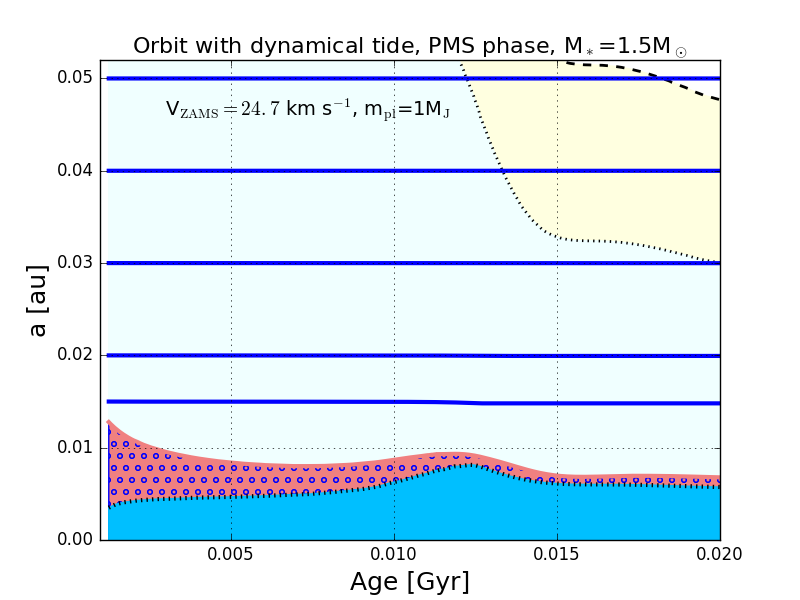}
\caption{{\it Left panel:} Same as the left panel of Fig.~\ref{P001z20S1PMS1aVSage1JUP}, but the mass of the star is 1.5 M$_\odot$. The blue continuous and dotted curves show the evolution of the orbits for planets
starting their evolution at various initial distances from their host star. The continuous curves show the orbital evolution for initial distances of 0.05, 0.04, 0.03, 0.02, and 0.015 au. The black dashed and dotted curves show $a_{\rm corot}$ and $a_{\rm min}$ for the case starting at 0.05 au, respectively. The horizontal dashed red line is the orbit starting at 0.03 au computed considering only the equilibrium tide. The pale blue region covers the zone below $a_{\rm min}$
(no dynamical tide active). The yellow region indicates where dynamical tides are active and tend to shrink the orbit (for the distance 0.05 au). The white area corresponds to zones where tides tend to widen the orbit (for the distance 0.05 au).
{\it Right panel:} Same as the left panel, but the 1.5 M$_\odot$ stellar model has a slower rotation.
}
\label{P1p5z14S51p0PMS}
\end{figure*}

We now consider a rotating 1.5 M$_\odot$ star with surface rotation during the PMS phase similar to that of the 1 M$_\odot$ model (see Fig.~\ref{struc}). 
Planet orbits of 1 Jupiter mass planets are shown in the left panel of Fig.~\ref{P1p5z14S51p0PMS}. The two main differences between the 1.5 M$_\odot$ model and the 1 M$_\odot$ are that on one hand,
the duration of the PMS phase is shorter by about a factor two. Second, in the case of the 1.5 M$_\odot$ star, the convective envelope disappears after about 12 Myr, which switches the tides off.
The main effect of these differences is that although tides are stronger when the mass of the star increases, they have less time to affect the orbit. As a result, the domain where an engulfment occurs
is smaller around the 1.5 M$_\odot$ star than around the 1M$_\odot$ star. For instance, in the case of a 1 Jupiter mass planet, no engulfment occurs. The reason is mainly that
for the closest planet (e.g., the planet beginning its evolution at 0.015au), the orbit is below $a_{\rm min}$ at the beginning and when the orbit crosses   $a_{\rm min}$, no outer convective zone exists any longer.
For a slightly larger initial distance (e.g., the case for 0.02 au), we observe a small drop when the orbit crosses $a_{\rm min}$. The drop is no longer observed when the orbit crosses $a_{\rm min}$ after the convective envelope 
has disappeared.

Interestingly, we note that the drop becomes larger when the initial distance increases, at least up to a point equal to $\sim$0.045 au. For these cases, the orbit is below $a_{\rm min}$ for a shorter time as the
distance increases, thus providing more time for the tide to shift the orbit down. Above 0.045 au, this trend is counterbalanced by two facts: the tidal torque decreases when the distance increases, and the orbit
crosses $a_{\rm corot}$ , causing the tides to widen the orbit from that point on.

Qualitatively, the behaviors are similar to those for the 1 M$_\odot$ star during the PMS phase, but the regions of engulfment and the planet desert become smaller.
This would be even more significant for cases with a delayed starting time.



\subsubsection{Effect of changing the stellar rotation}

In the right panel of Fig.~\ref{P1p5z14S51p0PMS}, we show the evolution of orbits for a Jupiter mass planet around a slow-rotating 1.5 M$_\odot$ stellar model. The situation is very different from the one obtained with the fast-rotating model: the orbits show hardly any change because of the slow rotation that pushes $a_{\rm min}$ outward  and thus implies that dynamical tides can be non-zero only at large distances, which are so large that
the amplitude of the tide becomes too small to significantly affect the orbit.
In this situation, only equilibrium tides are therefore important. These tides are too small to impact the orbits, however.
Only when the mass of the planet is higher (typically more than 13-15 Jupiter masses) can equilibrium tides affect the orbits of the closest planets.

This discussion illustrates that stellar rotation is a key factor for the evolution of the orbits during the PMS phase. Orbits are impacted by tides only around sufficiently fast rotating stars.

\subsection{Red giant phase}

The red giant phase differs from the PMS phase by its duration (about an order of magnitude longer), the radius of the star (which can be two orders of magnitude larger), and the very low value of the surface rotation
(this makes the factor $\epsilon$ in the dynamical tide expression very small). All these differences will favor equilibrium tides over dynamical tides during that phase.

\begin{figure*}
\centering
\includegraphics[width=.34\textwidth, angle=0]{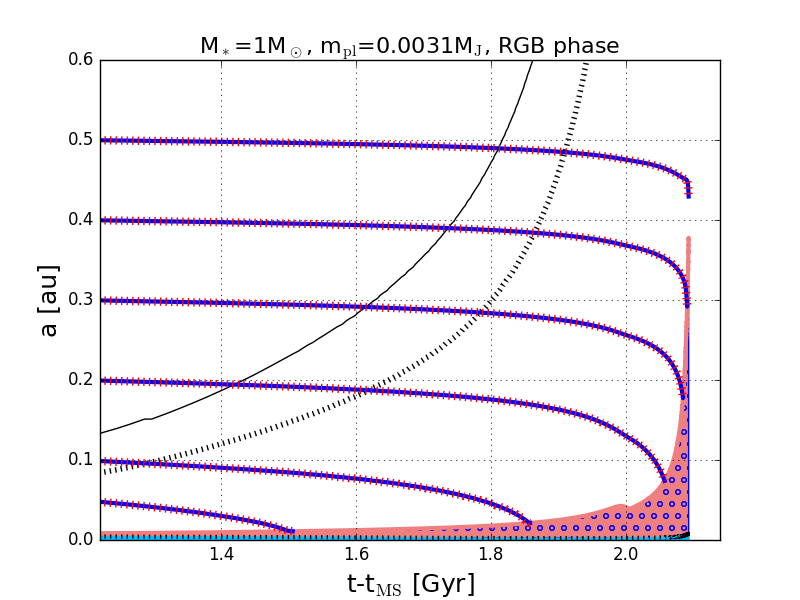}\includegraphics[width=.34\textwidth, angle=0]{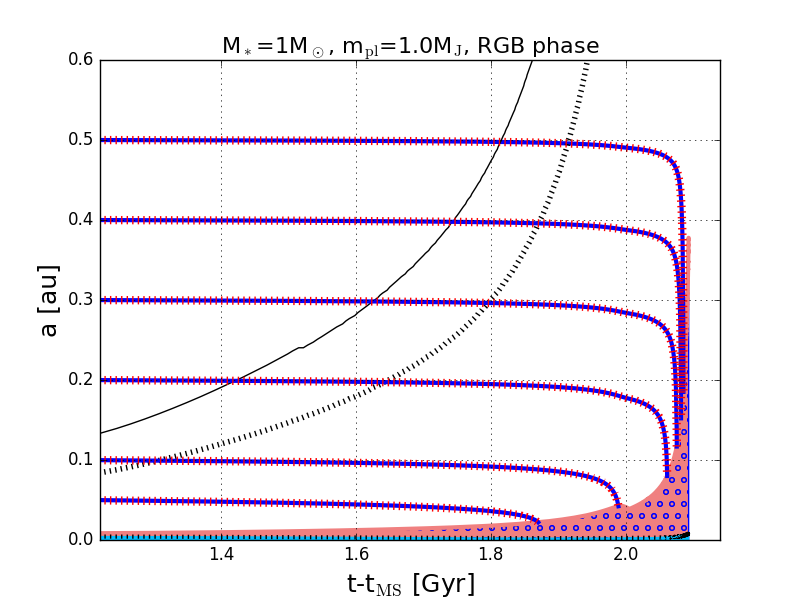}\includegraphics[width=.34\textwidth, angle=0]{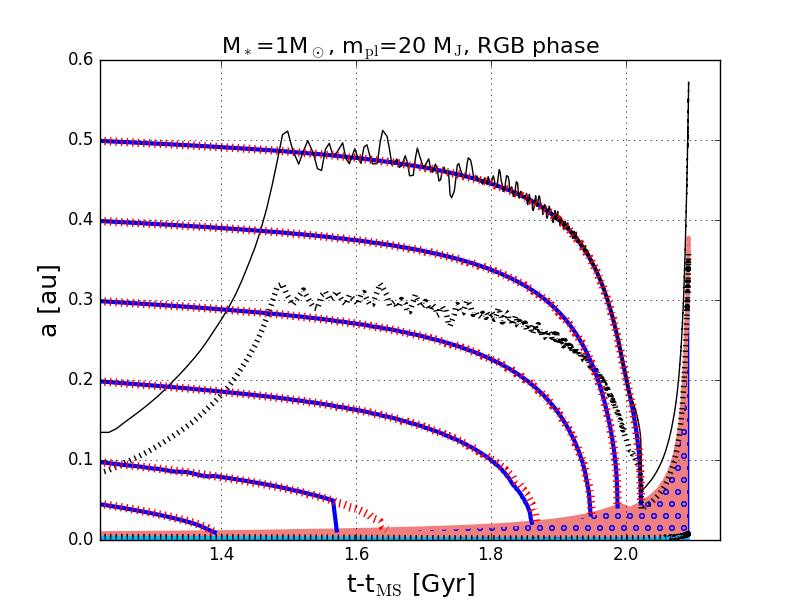}
\caption{Evolution of the orbits of 0.0031 (mass of the Earth), 1 and 20 Jupiter mass planets around a 1 M$_\odot$ star as a function of time $t$, where $t-t_{\rm MS}=0$ corresponds to the end of the main-sequence lifetime. The continuous and dotted black curves are $a_{\rm corot}$ and $a_{\rm min}$ , respectively, which correspond to the cases where the initial distance is 0.5 au. The blue curves are the orbits computed with both the equilibrium and dynamical tides. The red tick curves (nearly always superposed on the continuous blue lines) are the orbits computed without the dynamical tide (only equilibrium tide).}
\label{fig5}
\end{figure*}

\begin{figure*}
\centering
\includegraphics[width=.46\textwidth, angle=0]{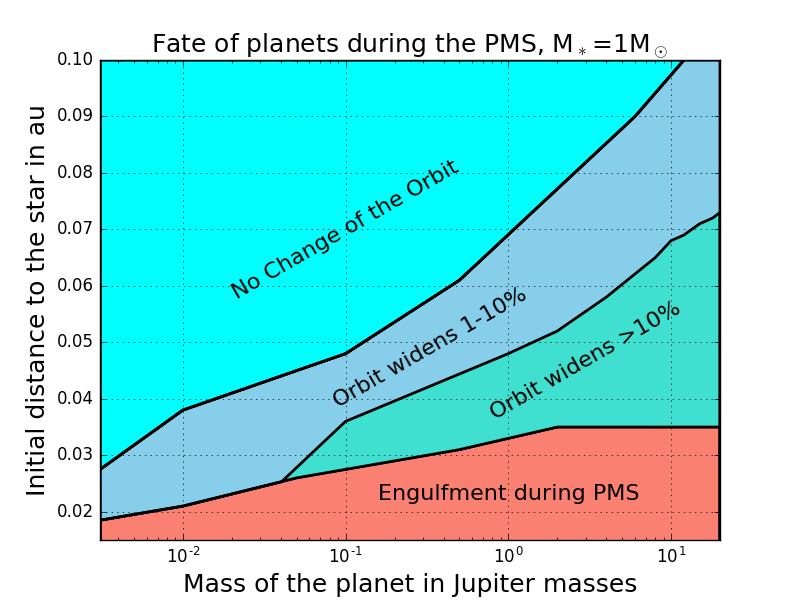}\includegraphics[width=.46\textwidth, angle=0]{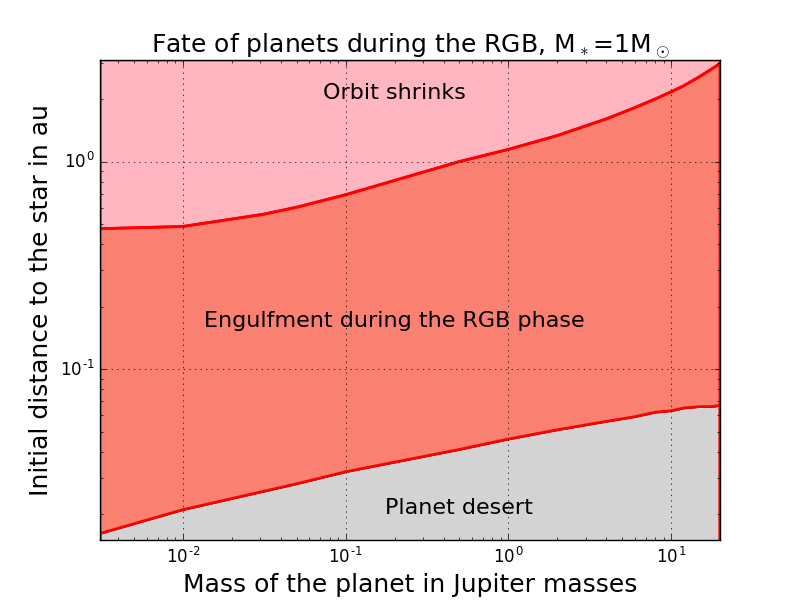}
\includegraphics[width=.46\textwidth, angle=0]{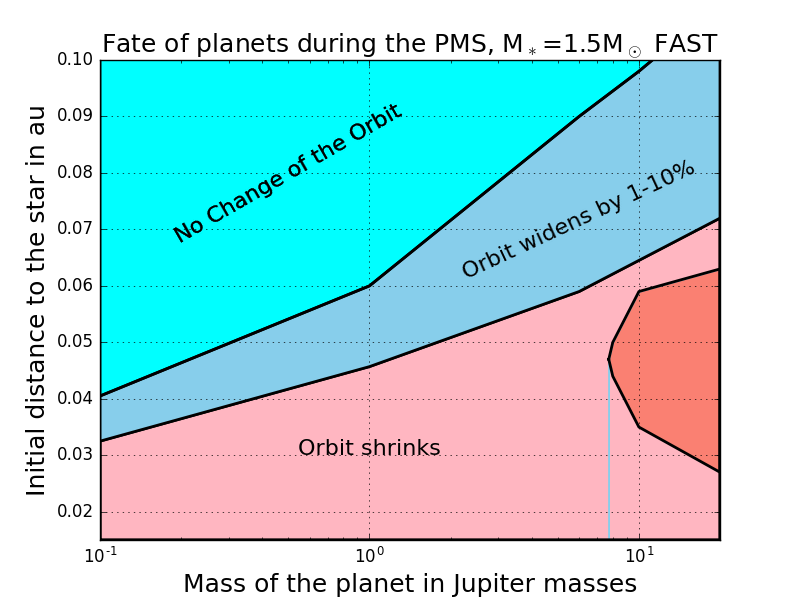}\includegraphics[width=.46\textwidth, angle=0]{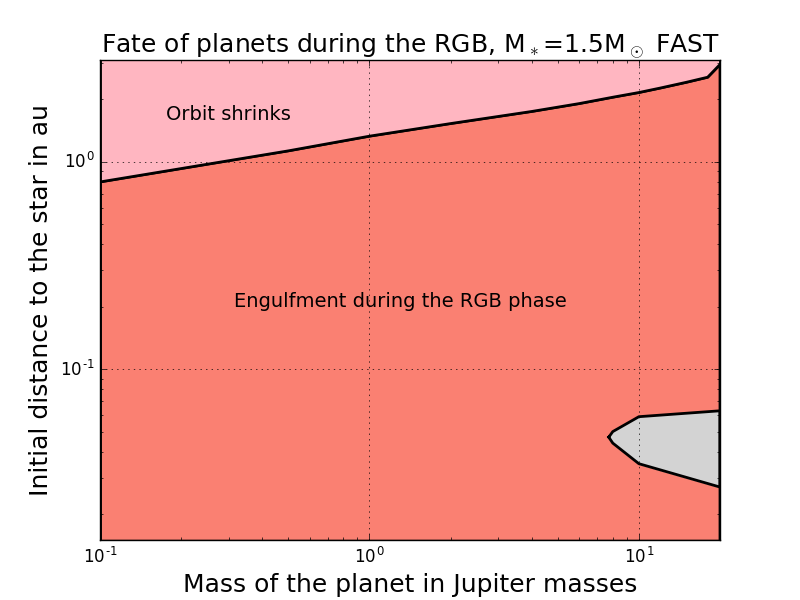}
\includegraphics[width=.46\textwidth, angle=0]{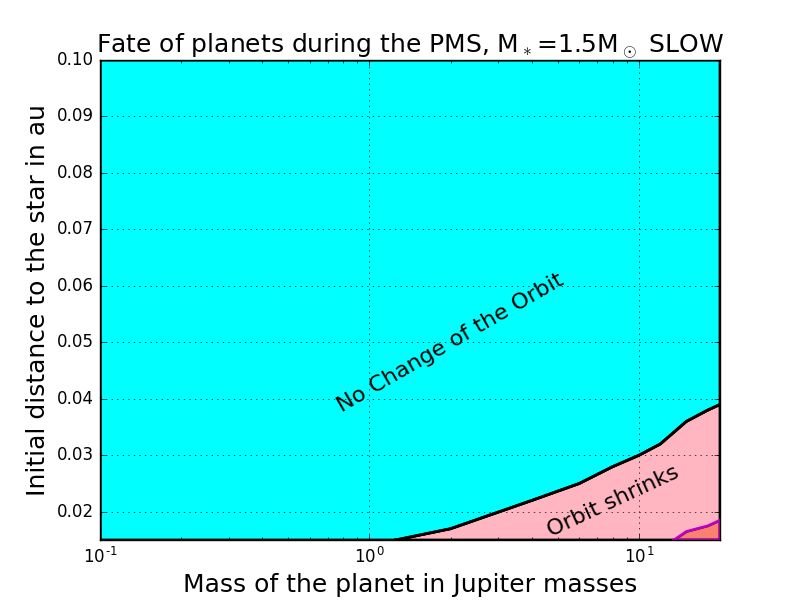}\includegraphics[width=.46\textwidth, angle=0]{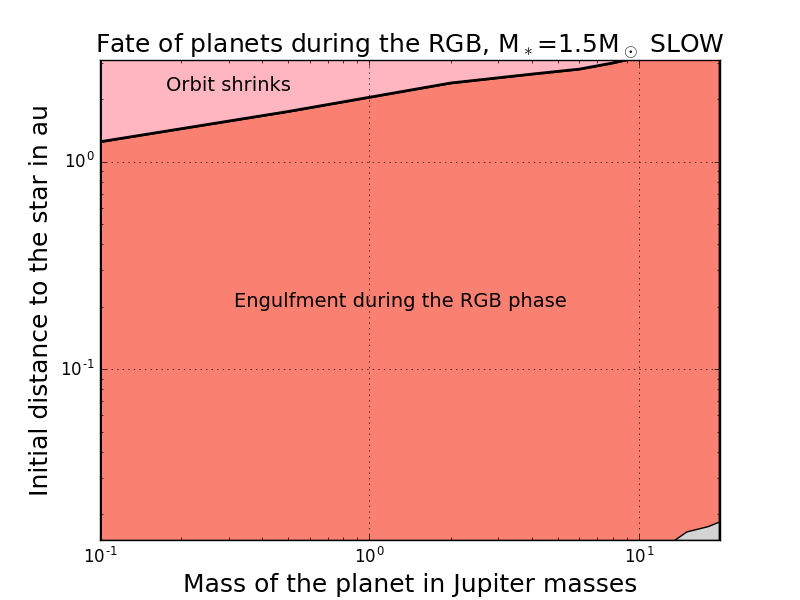}
\caption{Fate of planets of different masses (in Jupiter masses), starting their evolution at various distances (in au) from their host star. 
The left panels show the fate of planets during the PMS phase. The right panels show the fate of planets during the RGB phase.
The left panels cover a much wider range of distances (more than 10 times larger) than the left panel.
From top to bottom, the host star is a 1 M$_\odot$ star (see its surface rotation in Fig.~B1), a fast- and a slow-rotating 1.5 M$_\odot$ star, respectively. The domain of planet masses  considered for the 1.5 M$_\odot$ is smaller than the domain covered for the 1M$_\odot$ plot.
Planets in the upper blue regions show no change in their orbit.
Planets in the salmon region are engulfed.
The orbits of planets in skyblue and turquoise regions are widened by tides.
Orbits of planets in the pink regions were shrunken by tides.
The light gray zones correspond to the domains where no planet should be observed, either
because they have been engulfed during the PMS phase or because they have been kicked out from their original to a more distant orbit.
}
\label{FATES}
\end{figure*}

As expected from the simple considerations above, during the red giant phase, we obtain that equilibrium tides dominate the orbital evolution. 
An illustration of this is shown in Fig.~\ref{fig5}, where orbits of planets with masses equal to the Earth, 1 Jupiter mass, and 20 Jupiter masses
are shown as a function of time. For nearly all the cases considered here (there is only one exception), the orbits computed with and without the dynamical tides are identical. This reflects the fact that the equilibrium tides are the key factor
for the behavior shown. The exception is the 20 Jupiter mass planet at an initial distance of 0.1 au. When only equilibrium tides are accounted for, the engulfment occurs slightly later than when both tides work together to
shrink the orbit. The effect is very limited, however. Moreover, this evolution may also suffer from numerical limitations because of the interplay between evolution of the orbit and of $a_{\rm min}$ explained above. The bottom line of these comparisons is that equilibrium tides dominate the evolution of the orbits during the RGB phase.
Similar conclusions are obtained from orbit computations around the slow- and fast-rotating 1.5 M$_\odot$ models. This therefore indicates that the results obtained by 
\citet{GIOI, GIOII}, and \citet{GIOIII},  
where we studied the evolution of the orbits  and the impacts on the surface rotation  of the RGB stars for planets with masses between 1 and 15 Jupiter masses beginning their evolution with initial distances above 0.5 au, are unaffected by the fact that in these works the dynamical tides were neglected.






\section{Discussion of the fate of planets}\label{sec:3}

We have explored for each of our stellar models the fate of planets starting their evolution at different distances from their host stars. The results of this study are shown in Fig.~\ref{FATES}. We recall
that the limits shown in these figures depend on the detailed expressions of the tides, on the starting time (here we chose a starting time of 2 Myr and 1.2 Myr for the 1 and 1.5 M$_\odot$ stellar models, respectively), on the physics of the stellar models, and on the grid resolution (in terms of planet masses and distances) of the orbit computations. What we can hope to deduce here therefore are some trends that are robust enough to be independent of the above aspects.

\subsection{Pre-main-sequence phase}

The most striking feature concerning the fate of planets during the PMS phase is the sensitivity on the stellar rotation. This is well illustrated by comparing the middle left panel with the lower left panel of Fig.~\ref{FATES}.
When the rotation is slow, $a_{\rm min}$ is shifted outwards. This means that for the most close-in planets (where tides would be the most significant if active), dynamical tides have no chance to impact their orbit and the equilibrium tides are too small
and do not have enough time to have a strong impact. As described above, the domain where the orbits are affected by dynamical tides is therefore much smaller when the rotation is slow. This is illustrated here for a 1.5 M$_\odot$ stellar model,
but it would be the same for the 1 M$_\odot$ model if we were to chose lower rotation rates. This indicates that the distribution of the planets' distances to their host star during the main-sequence phase depends on the stellar rotation in the distance range
below about 0.1 au.


A comparison of the upper left plot with the middle left plot shows that the mass of the star affects the fate of the planets during the PMS phase. Increasing the planet mass,
on one hand, increases the tidal torque. On the other hand, increasing the stellar mass reduces the time during which the tidal torque can have an impact by reducing the time during which
a convective envelope is present. This explains why for planets with a mass lower than about 8 Jupiter masses, in the case of the 1.5 M$_\odot$, no engulfment is obtained down to a distance of 0.015 au.
Conditions for planet survival are therefore better around the 1.5 M$_\odot$ than around a 1 M$_\odot$ star with similar surface rotations.

\subsection{Red giant phase}

The situation during the RGB phase is simpler than during the PMS phase. This is in part because 
the behavior during that phase is entirely dominated by the equilibrium tide. 
The distance below which an engulfment is expected during the red giant branch $a_{\rm RGBeng}$ depends on the maximum radius reached
at the tip of the RGB phase and also on the duration of the RGB phase. The larger the maximum radius and the longer the RGB phase, the smaller $a_{\rm RGBeng}$.  This explains why 
$a_{\rm RGBeng}$ shifts downwards from the slow  to the fast 1.5 M$_\odot$ stellar models and then to the fast 1 M$_\odot$ stellar model.

Finally, we mention a byproduct of this study: According to the present computations, the orbit of an Earth-size planet at 1 au orbiting a 1 M$_\odot$ star would not  be affected during the PMS phase and would survive the
RGB phase.

\section{Discussion and conclusions}

One of the main aims of the present work was to determine the extent to which previous works focussing on the fate of planets during the RGB phase are affected by accounting for the dynamical tide.
Our conclusion is that dynamical tides in convective zones have little to no impact during that phase and thus do not affect these results. 

We have extended our previous works by studying what happens during the PMS phase and for much closer-in planets. We explored the fate of  planets orbiting their host stars at
distances as small as 0.015 au compared to 0.5 au in \citet{GIOI, GIOII,GIOIII}, and \citet{GIOIV}. The main points that we learned from our computations confirm previous works  \citep{Bol2016, Gal2017, Bol2017}, namely that
the fate of planets at distances below a few hundredths of 1 au (this limit depends on the mass of the planet) is mainly governed by dynamical tides. We showed the high sensitivity of the results
to the stellar rotation. A slow rotation quenches the activity of the dynamical tide and thus prevents any strong impact of tides on the planetary orbits. A slow rotation might be a consequence of planet formation.
Slow rotation might even be a consequence of a long
star-disk coupling phase during the PMS phase, and a long disk lifetime may favor planet formation. If true, 
this would limit the impact of dynamical tides in convective regions during the PMS phase.

On the other hand, when a planet orbit shrinks or when a planet is engulfed, the stellar rotation is accelerated, shifting $a_{\rm min}$ downwards (we recall here that for an orbital radius smaller than $a_{\rm min}$, the dynamical tide in the convective zone is zero)  and activating dynamical tides for still closer-in planets (which initially lay closer than $a_{\rm min}$). Moreover, $a_{\rm corot}$ (the corotation radius) is shifted downwards, enlarging the domain where planets are kicked out from their original orbits. It will certainly be interesting in the future to study the effect
of these processes not only on one planet, but in the frame of a planetary system. 


Our results suffer from some uncertainties, of course. 
An obvious uncertainty arises from the complexity of modeling the tidal interactions.
We did not account for the effects of dynamical tides in radiative regions \citep{Goodman1998, Ogilvie2007, Barker2010, Chernov2017, Weinberg2017}. 
In some circumstances (see below), including this type of tides may affect the fate of short-period planets with respect to 
what we obtained here, while in other circumstances, the effects may be small. We discuss the circumstances below when these tides probably have little effect and when
they might have a very significant effect.

 \citet{Ogilvie2007} indicate that Hough waves (i.e., those excited in radiative zones) are likely not dissipating energy in 
solar-type stars that are younger than or of the same age as the Sun, and that are hosts of hot Jupiters. Thus, in those situations, the inclusion of such tides is not expected to change our results.
On the other hand, these waves may be dissipative in stars older than the Sun and for short-planet periods (typically a few days, i.e., orbital distances of a few hundredths of 1 au). From this we conclude that the situation shown in the upper left panel of Fig.~6 will probably not be changed by the inclusion of the effects of dynamical tides in radiative zones, while it may change the results shown for distances smaller than 0.1 au
in the upper right panel. As a numerical estimate, \citet{Barker2010} predict that as a result of the action of the dynamical tide in radiative zones,  giant planets around G and K stars with orbital periods shorter than about two days might be engulfed. 
In that respect, it is interesting to mention the discussion of WASP-12 by \citet{Weinberg2017}, who proposed that the observed rate of decrease of the orbital period of this hot Jupiter orbiting a 1.2-1.3 M$_\odot$ (P$_{\rm orb}$=1.1 day) might be due to the action of dynamical tides in radiative zones provided the star is in its sub-giant phase. Indeed, for stars with a convective core, \citet{Barker2010} indicated that dynamical tides in radiative zones may be ineffective, thus dynamical tides in radiative zones are expected to be efficient
during the transition phase between the end of the main-sequence and the beginning of the convective He-burning core. This is also supported by the analysis of \citet{Chernov2017}. Inclusion of dynamical tides in radiative zones may therefore change the bottom part (for distances shorter than about 0.05 au) of the middle and bottom right panels of Fig.~6. 

As described in Sec. 2.2, the expression for the dynamical tide in the convective zone is obtained in the frame of a very schematic model for the star.
In particular, it is obtained
assuming that the star consists of two zones, a core and an envelope, each zone having a uniform density \citep{Ogil2013}. Actual convective zones have densities that can vary by orders of magnitudes. This stratification, if accounted for, might reduce the dissipation and thus might lead to less
efficient dynamical tides in convective zones.  Another weakness of the above expression is that it provides only a frequency-averaged dissipation rate. Actual dissipation rates at various frequencies may vary by several orders of magnitudes 
\citep[see, e.g., Fig.~6 in][]{Ogilvie2007}. This aspect adds a degree of uncertainty to the results obtained with the approach used here. On the other hand, works like ours, while still suffering from many uncertainties, allow the  physics of tides to be constrained
by providing some  theoretical predictions such as those shown in Fig.~6.

We considered here circular orbits in the stellar equatorial plane.  Accounting for the cases of eccentric and inclined orbits will certainly lead to significant changes in our mapping shown in Fig.~6, thus the limitations of the cases explored here have to be kept in mind.

Other uncertainties come from additional processes affecting the planets and their orbits.
We have accounted for the friction and gravitational drag. 
Frictional and gravitational drag forces tend to shrink the orbits and thus enlarge the zone of planet engulfment.  
In our computations, these terms have negligible impacts during the PMS phase, mainly because no stellar mass loss was considered and the number density of circumplanetary material was taken to be the value in the present-day solar system. This is likely not a very realistic assumption. A young star may lose some mass, and the circumplanetary material may have a density different from the density measured today in the solar system. 
During the RGB phase, these drag forces have non-negligible impacts and tend to shrink the orbit before tides become important. On their own, however, they are not strong enough
to produce an engulfment. Tides play the main role.

Other important effects on the orbit are the processes that change the mass of the planet. We here assumed constant-mass planets, but again this is not
realistic, since the planet can evaporate, or even accrete mass, or become disrupted by tides (if its orbital distance is below the Roche limit). 


Finally, we also mention that as a consequence of the evolution of the planetary orbits, some characteristics of the star will also change. 
We have discussed some of these points for what concerns the RGB phase in our previous works  \citep{GIOI, GIOII, GIOIII}, but a similar study must be made for the
PMS phase and the main-sequence phase. Regarding this, we can note that during the PMS phase, the convective zone recedes rapidly in mass when the star contracts toward the ZAMS. If during this
period the orbit of the planet shrinks, and if the angular momentum transferred from the orbit to the outer convective zone is large, it may produce a strong acceleration of the convective envelope of the star, which may have an impact
on mass-loss processes if the equatorial velocity of the star becomes equal or at least approaches the critical velocity (the velocity at which the centrifugal acceleration at the equator becomes
equal to the gravity). A work in preparation, accounting for the possibility of a change in the mass of the planet, currently studies whether such situations can occur.


\begin{acknowledgements} 
We thank the anonymous referee for the very valuable and important suggestions that allowed us to improve the paper. We thank our colleagues Florian Gallet and Eva Villaver for 
the very useful discussions of some aspects of this paper.
This project has been supported by the Swiss National Science Foundation grant 200020-172505.
\end{acknowledgements}

\begin{appendix}


\section{Values of the tidal dissipation for equilibrium tides}

The quantity $\sigma_\star$ that appears in Eq.~(3) can be written as \citep[see Eq. 113 in][]{Eggleton1998}
$$
\sigma_\star={2 \over M_\star^2 R_\star^4 Q_E^2} \int w l \gamma(r) {\rm d}m,
$$
where $Q_E$ is the normalized quadrupole moment, $w$ and $l$ are relevant velocities and lengths in the turbulent region, and $\gamma$
accounts for the tidally induced velocity field in the star (this $\gamma$ is not the same as the $\gamma$ appearing in Eqs. 6, 7, and 8.). The integral is taken over the turbulent regions of the star. 
\citet{Hansen2012}, replaced the product $wl$ in the integral, which is a viscosity, by the expression
$$
\nu_0 \upsilon_{\rm conv} H_{\rm p} f,
$$
where $\nu_0$ is a normalization constant, $ \upsilon_{\rm conv}$ is the convective velocity, $H_{\rm p}$ is the pressure scale height,
and $f$ accounts for the fact that the coupling between tides and the turbulent region may show a frequency dependence.
This term $f$ is the same as the term in Eq.~(1) of this paper.

The turnover time in a convective region is locally obtained by the ratio $H_{\rm p}/\upsilon_{\rm conv}$. Defining
$\tau$ as the average turnover time in the convective region defined by $1/M_{\rm env}\int (H_{\rm p}/\upsilon_{\rm conv}) {\rm d} m$ (where the integral is over the convective envelope), considering that $\gamma$ does not vary too much
in the whole convective envelope (see below), and taking for $H_{\rm p}$ a value equal to 0.05$R_\star$, we can write
$$
\sigma_\star \approx {0.005 \over M_\star^2 R_\star^2 Q_E^2} \nu_0   \gamma  {f \over \tau} M_{\rm env.}
$$
We can estimate $Q_E$ assuming that the star is nearly completely convective (polytrope with an index $n=1.5$) and using Eq.~(19) in 
\citet{Eggleton1998}. We obtain a value equal to 0.22. The value of $\gamma$ in case of polytrope with $n=1.5$ can be obtained
from the lower right panel of Fig.~1 in \citet{Eggleton1998}. An average value over the whole convective region is about 1, and
according to \citet{Hansen2012}, $ \nu_0$ is on the order of unity. Thus, we have
$$
\sigma_\star \approx {1 \over 10 M_\star R_\star^2} {f \over \tau} {M_{\rm env} \over M_\star}.
$$
This is very similar to Eq. (4). We also note that the value of 
3 10$^{-7}$ for  $\sigma_\star$
(normalized by 6.4 10$^{-59}$ g$^{-1}$ cm$^{-2}$ s$^{-1}$) 
given by \citet{Hansen2012} for a 1 M$_\odot$ at 1 Gyr with a forcing period of one day is similar to the value 1.8 10$^{-7}$ obtained by estimating, from our 1 M$_\odot$ model, the expression given by the right-hand term of Eq.~(4) for the same conditions. Given that the stellar models are not strictly the same, the agreement is reasonable. This shows the equivalence between the approach using $\sigma_\star$ and the approach using the dependence on the properties of the convective envelope as its mass and the turnover time. This latter approach is to be preferred over that of taking a constant $\sigma_\star$ when phases other than the main-sequence phase are considered and when studying the impact of the distance of the planet to its host star.


\section{Stellar quantities for computing the orbits}

\begin{figure*}
\centering
\includegraphics[width=.25\textwidth, angle=0]{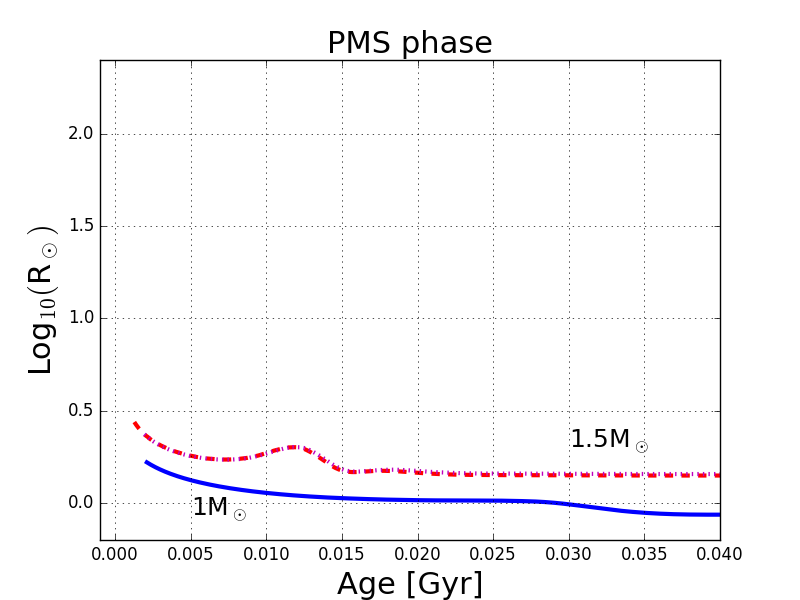}\includegraphics[width=.25\textwidth, angle=0]{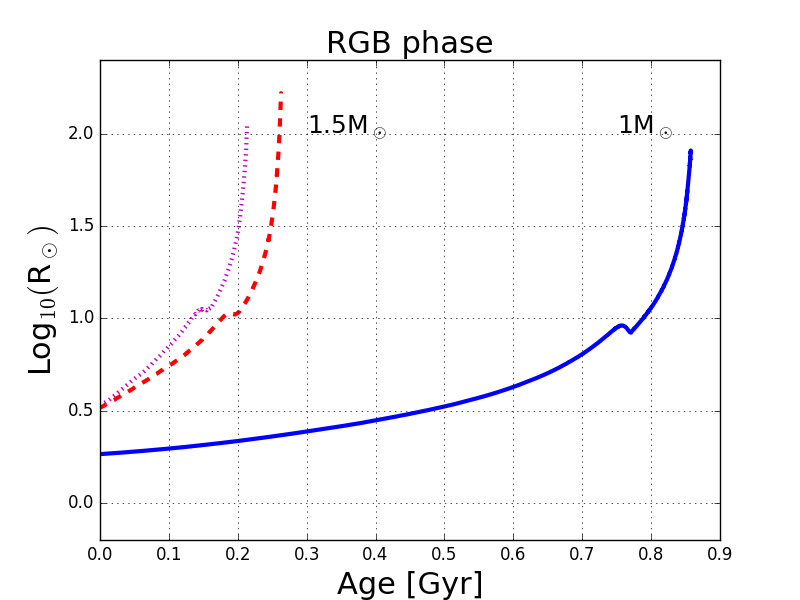}\includegraphics[width=.25\textwidth, angle=0]{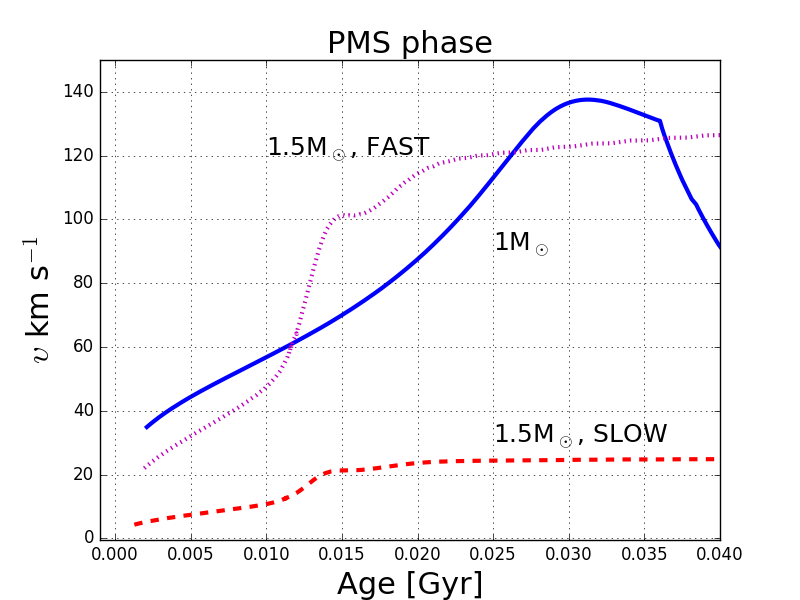}\includegraphics[width=.25\textwidth, angle=0]{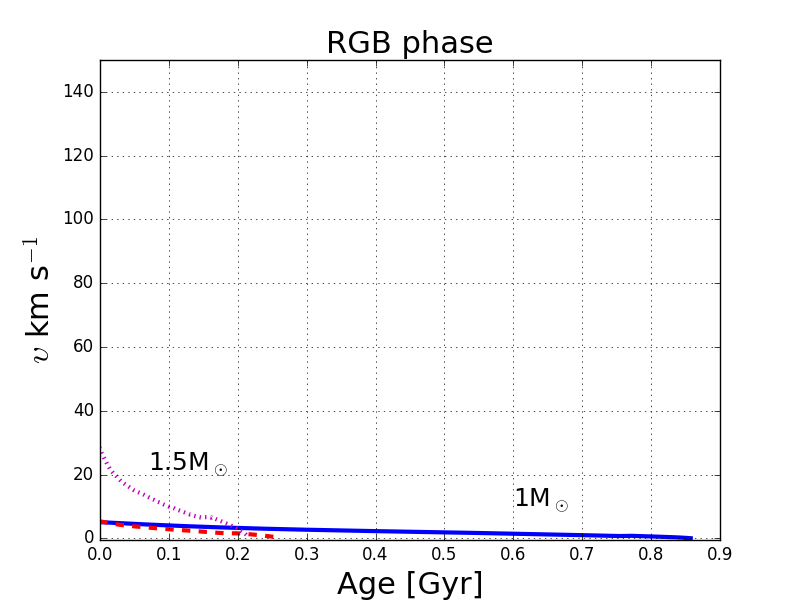}
\includegraphics[width=.25\textwidth, angle=0]{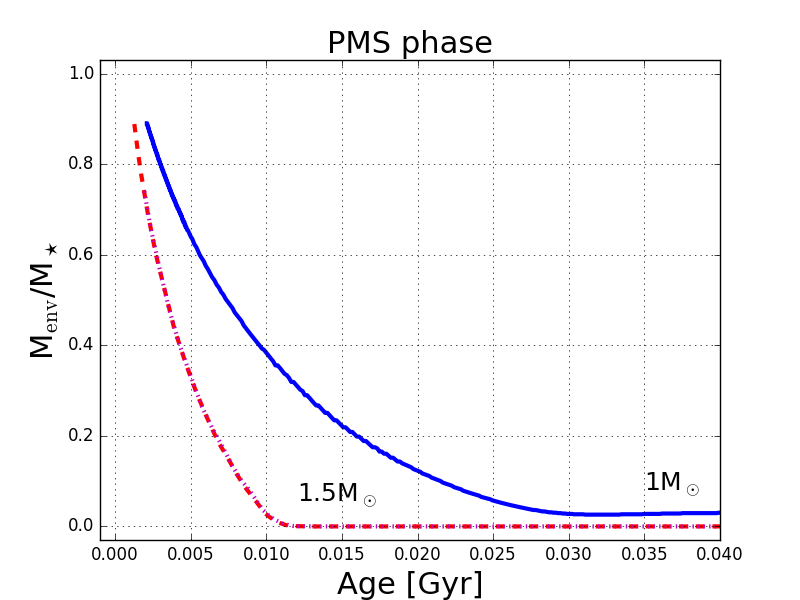}\includegraphics[width=.25\textwidth, angle=0]{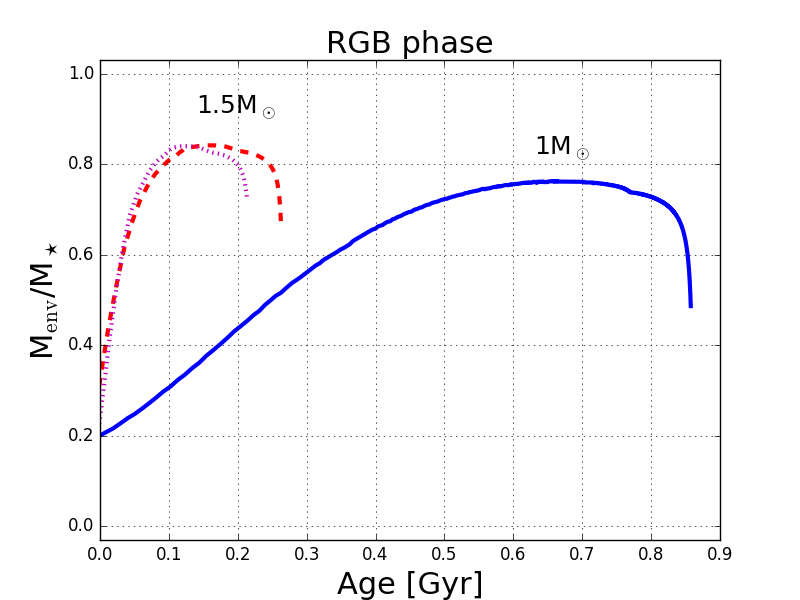}\includegraphics[width=.25\textwidth, angle=0]{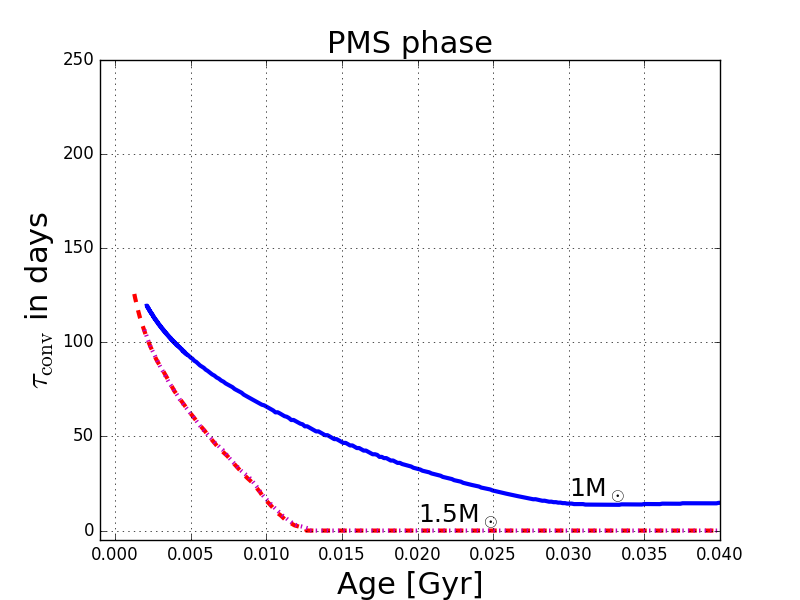}\includegraphics[width=.25\textwidth, angle=0]{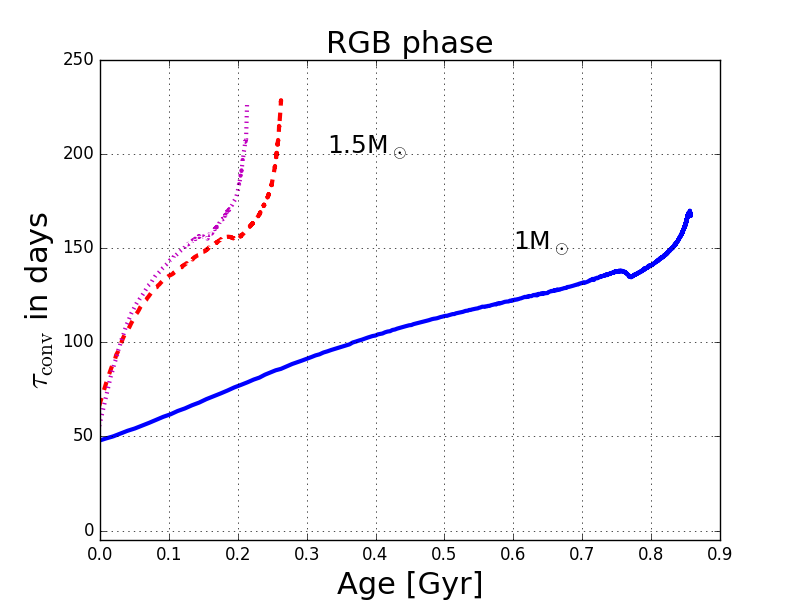}
\includegraphics[width=.25\textwidth, angle=0]{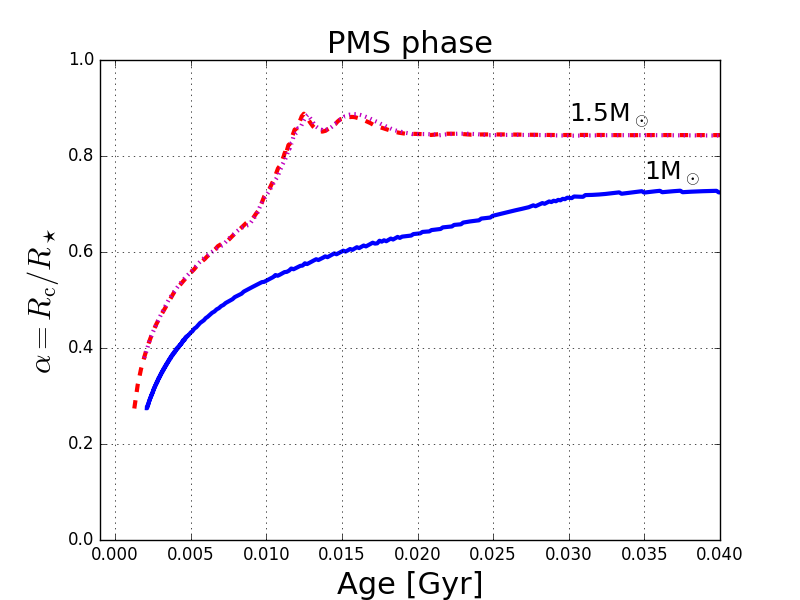}\includegraphics[width=.25\textwidth, angle=0]{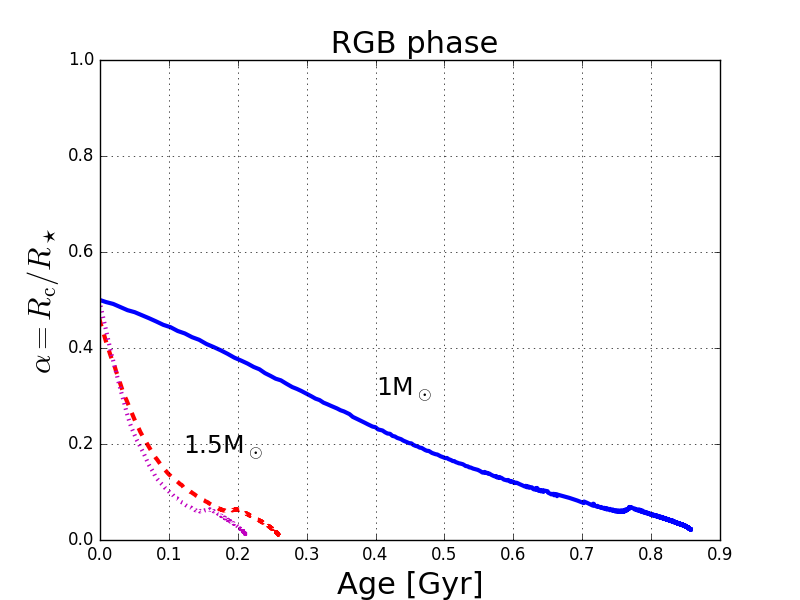}\includegraphics[width=.25\textwidth, angle=0]{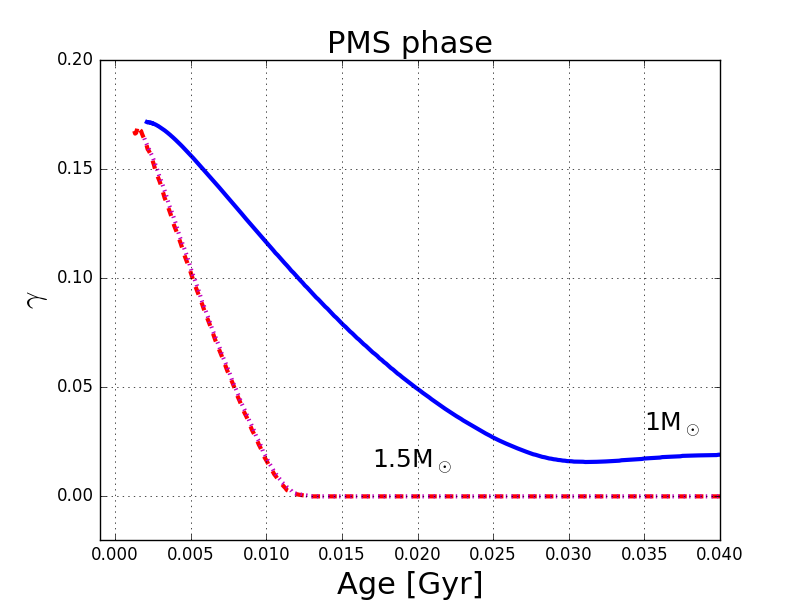}\includegraphics[width=.25\textwidth, angle=0]{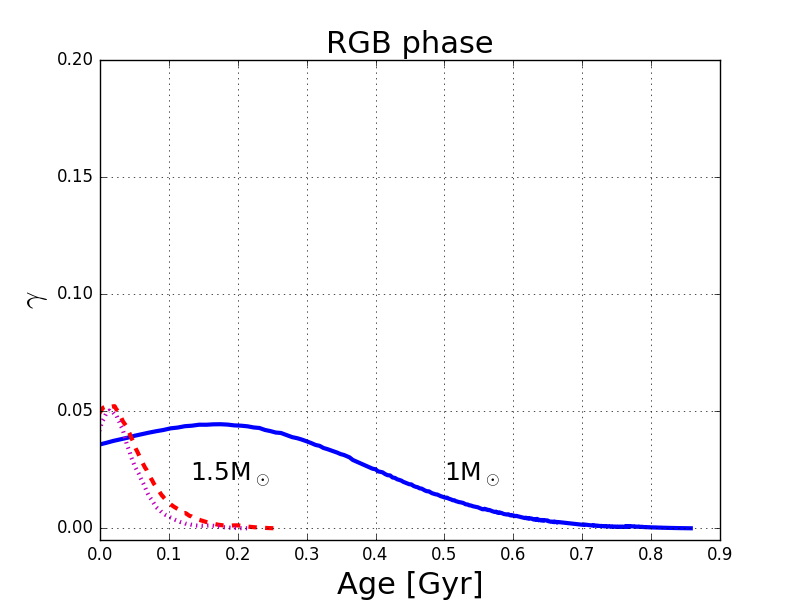}
\includegraphics[width=.25\textwidth, angle=0]{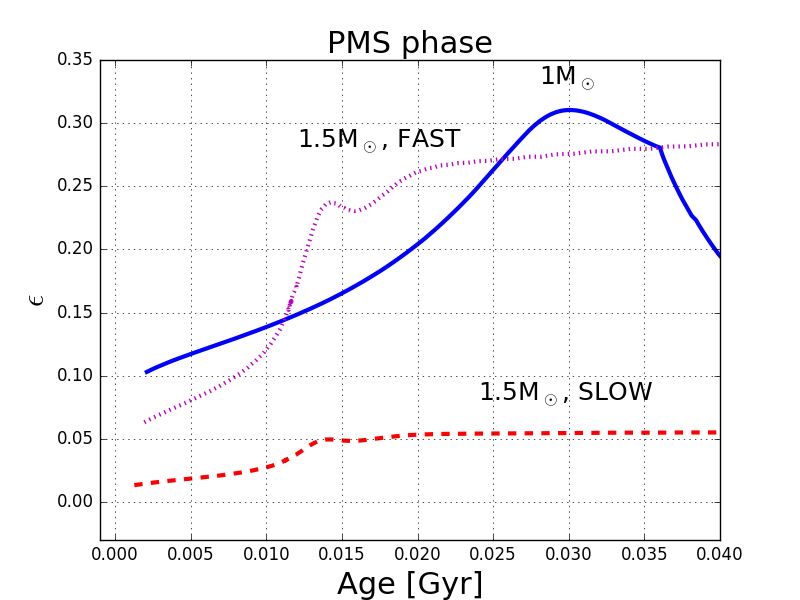}\includegraphics[width=.25\textwidth, angle=0]{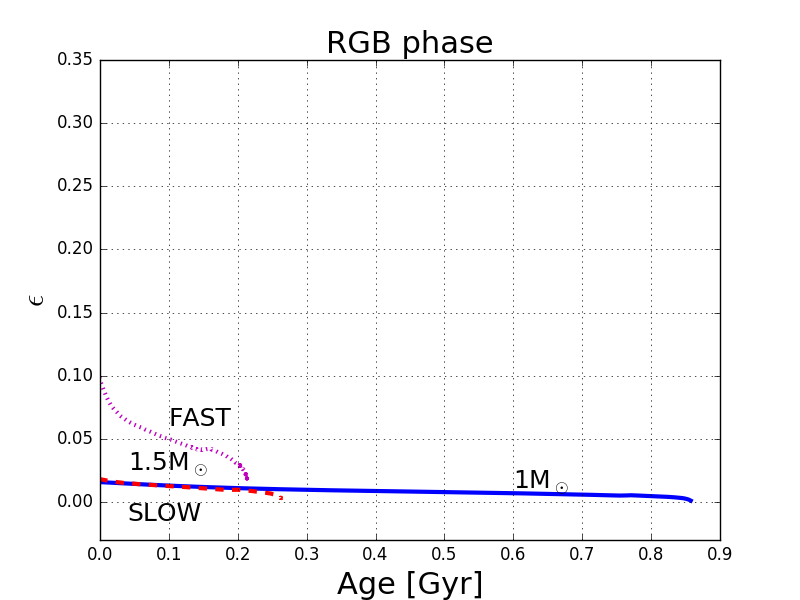}\includegraphics[width=.25\textwidth, angle=0]{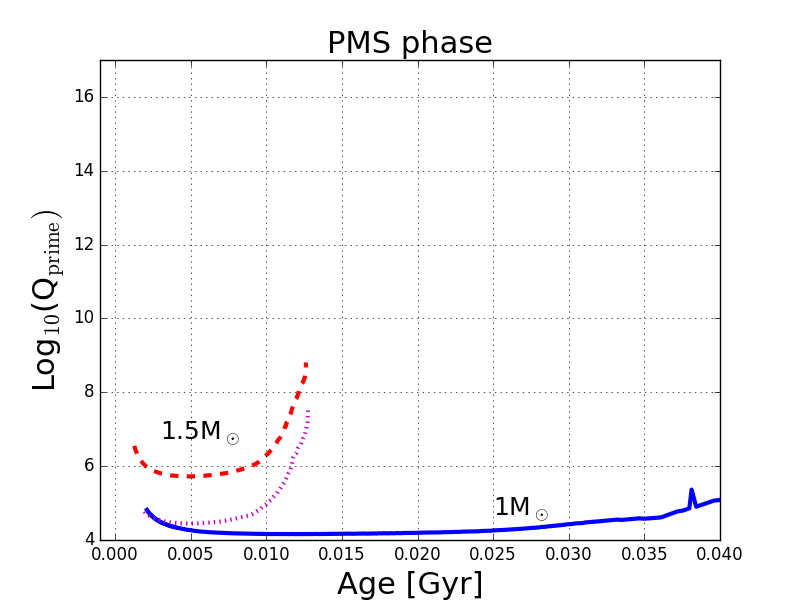}\includegraphics[width=.25\textwidth, angle=0]{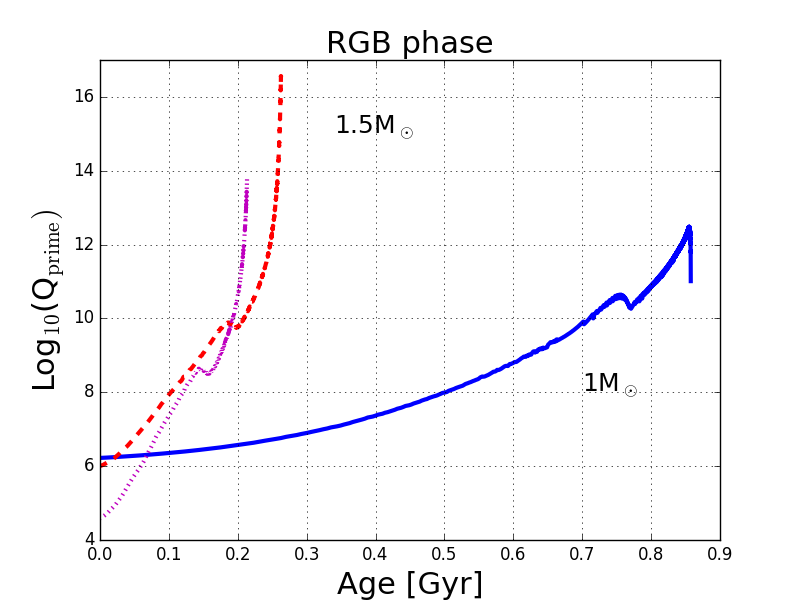}

\caption{Evolution as a function of time of various stellar quantities during the PMS and RGB phase for 1 (continuous blue line) and 1.5 M$_\odot$ stellar models. The 1.5 M$_\odot$ stellar model has been computed with two initial rotation velocities,
a slow (dashed red curves) and a fast velocity (dotted magenta curves).  
For the RGB phase, time equal to 0 corresponds to the local minimum of the luminosity before the
ascent of the RGB. The first two columns of plots show (from top to bottom) the evolution of the stellar radius, of the mass fraction of the convective envelope, of the ratio between the radius of the stellar core (zone below the convective envelope), and the total radius and the ratio between the surface angular velocity and the Keplerian angular velocity. The last two columns show (from top to bottom) the evolution of the surface equatorial velocities, of the convective turnover time, of $\gamma$ (the one appearing in Eqs. 6, 7, and 8), and Q$_{\rm prime}$, quantities needed for computing the dynamical tide. For comparison purposes, the range covered by the vertical axis is the same for the PMS and RGB phase.
} 
\label{struc}
\end{figure*}

Figure~\ref{struc} compares the evolution of various stellar quantities during the PMS and RGB phases of 1 and 1.5 M$_\odot$ models. For the purpose of comparisons, the range of the vertical axis is taken to be the same for both phases. The quantities that differ the most between the PMS and the RGB phases are listed below.\begin{enumerate}
\item {\bf The duration.}  Typically, the time for the star to evolve from the Hayashi track to a time when typically a mass fraction of three thousands of hydrogen has been transformed into helium is about 70 Myr for the 1 M$_\odot$ model and 
30 Myr for the 1.5 M$_\odot$ (the rotation has little impact on that quantity). In the case of 1.5 M$_\odot$, the duration of the phase during which an outer convective zone is present is only about 12 Myr.
The duration of the ascent of the RGB is 850 Myr for the 1 M$_\odot$ and 200-270 Myr (depending on rotation) for the 1.5 M$_\odot$ model.  Thus the RGB phase is more than an order of magnitude longer
than the PMS phase.
\item {\bf The radius.}  During the PMS phase, the radius is a few solar radii, while it reaches values of up to  250 R$_\odot$ during the RGB phase. This is an increase by up to two orders of magnitude.
\item {\bf The velocity.} Velocities during the PMS depend on the value chosen as the initial value and are therefore high when the initial value is high and low otherwise. During the RGB phase, velocities are modest even
when a high initial rotation has been chosen at the beginning \citep[see also][]{GIOII}. This also implies that the quantity $\epsilon,$ that is, the ratio between the actual angular surface velocity and the
Keplerian angular velocity, is also smaller during the RGB than during the PMS phase.
\item {\bf Q$_{\rm prime}$.} This quantity is larger  during the RGB than during the PMS, which means that the dynamical tide is much less important during the RGB phase than during the PMS phase.
\end{enumerate}

\end{appendix}

\bibliographystyle{aa} 
\bibliography{biblio} 

\end{document}